\documentclass{new_tlp}

\usepackage[utf8]{inputenc} 
\usepackage[T1]{fontenc}

\usepackage[ruled,vlined]{algorithm2e}
\SetArgSty{textnormal} 

\usepackage{amssymb}
\usepackage{amsmath}
\usepackage{textcomp}
\usepackage{calc}
\usepackage{comment}
\usepackage[justification=centering]{caption}
\usepackage{subcaption}
\usepackage[hidelinks,pdftex]{hyperref}
\usepackage{relsize}

\usepackage{mathptmx}
\usepackage{dsfont}

\usepackage{calrsfs}
\DeclareMathAlphabet{\pazocal}{OMS}{zplm}{m}{n}

\usepackage{xspace}  

\usepackage{enumitem}
\setlist[enumerate,2]{ref=\arabic{enumi}.\alph*}

\usepackage{multirow}
\usepackage{longtable}
\usepackage{booktabs}

\usepackage{adjustbox}
\usepackage{rotating}

\usepackage{relsize}

\usepackage{url}

\usepackage{graphicx} 

\usepackage{xcolor}
\definecolor{PrologPredicate}{RGB}{0,0,200}
\definecolor{PrologVar}      {RGB}{145,032,039}
\definecolor{PrologComment}  {RGB}{169,082,044}
\definecolor{PrologOther}    {rgb}{0.2,0.2,0.2}
\definecolor{PrologString}   {RGB}{070,120,200}

\usepackage{listings} 
\usepackage{textcomp}
\newcommand{\code}{\lstinline[style=MyInline]}
\lstdefinestyle{tree}
{
  basicstyle = \small\ttfamily\color{PrologPredicate},
  basewidth = 0.5em,
  moredelim = {[s][\color{PrologString}]{ \{}{\} }},
  moredelim = {*[s][{\color{PrologVar}}]{(}{)}},
  literate     =
  {.\\=.}{{\ \char"5C=\ }}3
  {\\=}{{\ \char"5C=\ }}3
  {.<.}{{\ \#<\ }}4
  {.>.}{{\ \#>\ }}4
  {=}{{\ =\ }}3
  {.=.}{{\ \#=\ }}3
  {.=<.}{{\ \#=<\ }}5
  {.>=.}{{\ \#>=\ }}5
}
\lstdefinestyle{MyInline}
{
  basicstyle = \ttfamily\color{PrologOther},
  breaklines = true,
  breakatwhitespace=true,
  upquote = true,
  literate =
  {,}{}{0\discretionary{,}{}{,}}
  {|}{{\color{PrologOther}$\mid$}}1
  {\\\{}{{\color{PrologOther}\{}}1
  {\\\}}{{\color{PrologOther}\}}}1
  {[}{{\color{PrologOther}\small[}}1
  {]}{{\color{PrologOther}\small]}}1
  {.=.}{{\color{PrologOther}\#=}}3
  {.<.}{{\color{PrologOther}\#<}}3
  {.>.}{{\color{PrologOther}\#>}}3
  {.=<.}{{\color{PrologOther}\#=<}}4
  {.>=.}{{\color{PrologOther}\#>=}}4
  {<}{{\color{PrologOther}<}}2
  {>}{{\color{PrologOther}>}}2
  {=<}{{\color{PrologOther}=<}}3
  {>=}{{\color{PrologOther}>}}3
  {.\\=.}{{\color{PrologOther}{\char"5C}=}}3
  {\\=}{{\color{PrologOther}\char"5C=}}3
  {?-}{{\color{PrologOther}?-\,}}3
  {:-}{{\color{PrologOther}:-\,}}3
  {,}{{\color{PrologOther}\footnotesize,}}1  
  {\\$}{{\$}}1
}
\lstdefinestyle{MyProlog}
{
  keywords = {},
  upquote = true,
  basicstyle = \relsize{0}\ttfamily\color{PrologPredicate},
  basewidth = 0.48em,
  moredelim = {**[s][\color{PrologString}]{'}{'}},
  moredelim = {**[is][\color{PrologComment}]{`}{`}},
  moredelim = {**[is][\color{PrologPredicate}]{@}{@}},
  moredelim = {*[s][\color{PrologVar}]{(}{)}},
  moredelim = {*[s][\color{PrologOther}]{:-}{.}},
  moredelim = {*[s][\color{red}]{/*}{*/}},
  commentstyle = \mdseries\color{PrologComment},
  morecomment=[l]\%,
  morecomment=[s]{/*}{*/},
  literate     =
  {|}{{\color{PrologOther}$\mid$}}1
  {[}{{\color{PrologOther}\small[}}1
  {]}{{\color{PrologOther}\small]}}1
  {\\$}{{\$}}1
  {&(}{{\color{PrologOther}(}}1
  {&)}{{\color{PrologOther})}}1
  {&.}{{.}}0
  {.=.}{{\color{PrologOther}\ \#=\ }}3
  {.<.}{{\color{PrologOther}\ \#<\ }}3
  {.>.}{{\color{PrologOther}\ \#>\ }}3
  {.=<.}{{\color{PrologOther}\ \#=<\ }}4
  {.>=.}{{\color{PrologOther}\ \#>=\ }}4
  {.\\=.}{{\color{PrologOther}\char"5C=}}3
  {\\=}{{\color{PrologOther}\char"5C=}}3
  {,}{{\color{PrologOther}\footnotesize,}}1
  {;}{{\color{PrologOther}\footnotesize;}}1,
}

\lstdefinestyle{MyASP}
{
  keywords = {},
  upquote = true,
  basicstyle = \small\ttfamily\color{PrologPredicate},
  basewidth = 0.52em,
  moredelim = {*[s][\color{PrologString}]{'}{'}},
  moredelim = {*[s][\color{PrologString}]{"}{"}},
  moredelim = {*[s][\color{PrologVar}]{(}{)}},
  moredelim = {*[s][\color{PrologOther}]{:-}{.}},
  commentstyle = \color{PrologComment},
  morecomment=[l]\%,
  morecomment=[s]{/*}{*/},
  morecomment=[l]?,
  literate     =
  {..}{..}2
  {&(}{{\color{PrologOther}(}}1
  {&)}{{\color{PrologOther})}}1
  {&.}{{.}}0
  {,}{{\footnotesize,}}1
}

\usepackage{tikz}
\usetikzlibrary{shapes.geometric, arrows, patterns}

\usepackage[edges]{forest}

\usepackage{morefloats}
\usepackage{xcolor}
\usepackage[textwidth=0.2\textwidth,textsize=footnotesize]{todonotes}
\let\proof\relax

\usepackage{amsthm}

\let\oldc\,
\renewcommand{\,}{\oldc \allowbreak}
\let\oldland\land
\renewcommand{\land}{\oldland \allowbreak}

\newcommand{\goaltt}[2]{\ensuremath{\langle \mathtt{#1},\, \mathtt{#2} \rangle}}
\newcommand{\mt}[1]{\ensuremath{\mathtt{#1}}}

\newcommand*{\pr}[1]{%
  \ifcase#1%
  \relax
  \or \ensuremath{\mathtt{_1}}\nobreak
  \or \ensuremath{\mathtt{_2}}\nobreak
  \or \ensuremath{\mathtt{_3}}\nobreak
  \or \ensuremath{\mathtt{_4}}\nobreak
  \or \ensuremath{\mathtt{_5}}\nobreak
  \else
  \fi
}

\newcommand{\papermaterial}{\url{http://www.cliplab.org/papers/tclp-plai-iclp2019}}

\title[Abstract Interpretation Fixpoint using TCLP]%
{Evaluation of the Implementation of an Abstract Interpretation Algorithm using Tabled CLP \thanks{ Work partially supported by EIT
    Digital (https://eitdigital.eu), MINECO project
    TIN2015-67522-C3-1-R (TRACES), and Comunidad de Madrid project
    S2018/TCS-4339 BLOQUES-CM co-funded by EIE Funds of the European
    Union.
  } }

\author[J. Arias and M. Carro]{ 
JOAQU\'IN ARIAS and MANUEL CARRO \\
IMDEA Software Institute and Universidad Politécnica de Madrid\\
\texttt{\emph{joaquin.arias@\{imdea.org,alumnos.upm.es\}, manuel.carro@\{imdea.org,upm.es\}}} 
}

\volume{\textbf{10} (3):}
\jdate{March 2002}
\pubyear{2002}

\newcommand{\mtclp}{\mbox{Mod TCLP}\xspace}
\newcommand{\atclp}{\mbox{Aggregate-TCLP}\xspace}
\newcommand{\gtclp}{\mbox{TCLP}\xspace}

\begin{document}

\maketitle

\begin{abstract}
  \label{sec:abstract}

  CiaoPP is an analyzer and optimizer for logic programs, part of the
  Ciao Prolog system.  It includes PLAI, a fixpoint algorithm for the
  abstract interpretation of logic programs which we adapt to use
  \emph{tabled constraint logic programming}.  In this adaptation, the
  tabling engine drives the fixpoint computation, while the constraint
  solver handles the LUB of the abstract substitutions of different
  clauses.  That simplifies the code and improves performance, since
  termination, dependencies, and some crucial operations (e.g., branch
  switching and resumption) are directly handled by the tabling
  engine.  Determining whether the fixpoint has been reached uses
  \emph{semantic equivalence}, which can decide that two syntactically
  different abstract substitutions represent the same element in the
  abstract domain.  Therefore, the tabling analyzer can reuse answers
  in more cases than an analyzer using syntactical equality.  This
  helps achieve better performance, even taking into account the
  additional cost associated to these checks.  Our implementation is
  based on the TCLP framework available in Ciao Prolog and is
  one-third the size of the initial fixpoint implementation in CiaoPP.
  Its performance has been evaluated by analyzing several programs
  using different abstract domains.

  \smallskip\noindent
  This paper is under consideration for publication in Theory and
  Practice of Logic Programming (TPLP).
\end{abstract}

\begin{keywords}
   Abstract Interpretation, Constraints, Tabling, Prolog, PLAI.
\end{keywords}

\section{Introduction}
\label{sec:introduction}

Tabling~\cite{tamaki.iclp86,Warren92} is an execution strategy for
logic programs that suspends repeated calls which could cause infinite
loops.  Answers from non-looping branches are used to resume suspended
calls which can, in turn, generate more answers and resume other
suspended calls.  Only new answers are saved, and evaluation finishes
when no new answers can be generated.  Tabled evaluation always
terminates for calls/programs with the bounded term depth property
(i.e., they can only generate terms with a fixed finite depth) and can
improve efficiency for terminating programs which repeat computations,
as it automatically implements a variant of dynamic
programming. Tabling has been successfully applied in a variety of
contexts, including deductive databases, program analysis, semantic
Web reasoning, and model checking.

Constraint Logic Programming (CLP)~\cite{survey94} extends Logic
Programming (LP) with variables that can belong to arbitrary
constraint domains and the ability to incrementally solve equations
involving these variables.  CLP brings additional expressive power to
LP, since constraints can very concisely capture complex
relationships.  Also, shifting from ``generate-and-test'' to
``constraint-and-generate'' patterns reduces the search tree and
therefore brings additional performance, even if constraint solving is
in general more expensive than unification.

The integration of tabling and constraint solvers makes it possible to
exploit their synergy in several application fields: abstract
interpretation~\cite{swift2010:subsumption}, reasoning on ontologies,
and constraint-based verification~\cite{navas2013-FTCLP}.
In this paper we use \mtclp~\cite{TCLP-tplp2019} to adapt PLAI, the
fixpoint algorithm implemented in the program analysis, optimization,
and transformation tool
CiaoPP~\cite{hermenegildo11:ciao-design-tplp,ciaopp-sas03-journal-scp}.
The re-implementation of PLAI uses tabling to reach the fixpoint
(following ideas similar
to~\cite{kanamori93:absint-oldt,janssens98:tabling_for_AI-tapd}),
incremental aggregation
techniques~\cite{tabling-modes-GuoG08,zhou2010mode,swift2010:subsumption,atclp-padl2019}
to join the answers, by discarding the more particular ones,
and call entailment
checks~\cite{chico-tclp-flops2012-short,TCLP-tplp2019} to detect
repeated calls (in order to suspend execution to reuse answers from
previous calls), thereby speeding up convergence. The resulting code
space is reduced to one third and, consequently, increases the
maintainability of the abstract interpreter.

\section{Related Work}
\label{sec:related-work}

Abstract interpretation has always been seen as one of the most clear
applications of tabled logic programming.  It requires a fixpoint
procedure, often implemented using memo tables and dependency
tracking, which play a role very similar to the internal data
structures that tabling engines need to detect repeated calls,
store and reuse answers, and check for termination.


The relationship between abstract interpretation and tabling was
recognized very early.  \emph{Extension tables}~\cite{dietrich87} were
proposed to record results from the execution of predicates and turn
intensional definitions into extensional definitions.  Their
applications included ``improving the termination and completeness
characteristics of depth-first evaluation strategies in the presence
of recursion''.  The idea of extension tables were applied as the
embryo of SLG resolution and the XSB system.  At the same time,
abstract interpretation was then viewed as inefficient, and as part of
the efforts to make it a practical technique to implement analyzers,
tables, but also other ideas such as dependency tracking, were
used~\cite{pracabsin}, thus making it clear that a common underlying
technology could be used in both types of systems.

The next step was to use these components, independently available in
tabling systems, to explore how they could be used to build abstract
interpreters.  Earlier work~\cite{kanamori93:absint-oldt} explored the
possibilities offered by OLDT~\cite{tamaki.iclp86} to implement
abstract interpretation.  Using type inference as the guiding example,
it suggests certain changes to OLDT and concludes that it is feasible
to do abstract interpretation with OLDT.  The paper neither describes
an implementation nor reports performance, but it states that the
abstract interpreter was implemented and was available.
In~\cite{warren:tabled_prolog} an abstract interpreter written in XSB
is presented as one of the applications of tabled Prolog.

However, surprisingly few examples of abstract interpreters
implemented using tabling have been presented and evaluated w.r.t.\
implementations without tabling.  One of them
is a framework~\cite{janssens98:tabling_for_AI-tapd} based on
abstract compilation
that executes the abstract version of the program under analysis,
together with domain-dependent abstract operations, which is evaluated using
the tabling system XSB 
and compared with the AMAI and PLAI
systems~\cite{janssens1995blueprint,ai-jlp-short}. Both systems use
abstract interpreters written in Prolog without tabling, but they rely
on very different underlying technologies, and with different
representations for the abstract domains.
From that evaluation, the paper concludes that tabling is a viable
infrastructure for abstract interpretation, but concedes that the
PLAI fixpoint algorithm  was the most
efficient abstract interpreter for logic programming available at the
moment.  The very different underlying infrastructure makes it
difficult to use these results to draw meaningful conclusions.

On the other hand, abstract interpretation has been used
as a benchmark to compare different implementations and/or scheduling strategies of
tabling~\cite{demoen98:cat,freire01:_beyon_depth_first}.  Advanced
tabled systems and techniques have been proposed to implement more
efficient abstract interpreters by using the \emph{least upper bound}
operator~\cite{tchr2007} to combine answers, numeric constraint
solvers~\cite{chico-tclp-flops2012-short} to implement the Octagon
domain, and the \emph{partial order answer subsumption with
  abstraction}~\cite{swift2010:subsumption} for cases where, e.g., the
program computed does not have a finite model. However, none of them
reports performance evaluation against other frameworks.

In this paper we started with PLAI, the state-of-the-art abstract
interpreter used by CiaoPP, and re-implemented its fixpoint procedure
in Tabled CLP preserving the interface with the rest of the system.
Therefore, we can compare some indicators of code complexity (e.g.,
comparing lines of code, with the assumption that the tabled version
is essentially a subset of the original version) and performance on a
completely equal footing. This is, to our best knowledge, the first
comparison that has these characteristics.

\section{Background}
\label{sec:background}

\renewcommand{\u}{\vec{u}}
\renewcommand{\l}{\lambda}
\renewcommand{\b}{\beta}

In this section we briefly describe 
\mtclp~\cite{TCLP-tplp2019}, a generic interface that facilitates the
integration of constraint solvers with the tabling engine in Ciao,
\atclp~\cite{atclp-padl2019}, a framework implemented on top of \mtclp
to incrementally compute lattice-based aggregates,
and PLAI, the fixpoint algorithm used by CiaoPP.

\subsection{The \mtclp framework}
\label{sec:modul-tclp-fram}

Tabled Logic Programming with Constraints
(TCLP)~\cite{TCLP-tplp2019,tchr2007,bao2000} improves program
expressiveness and, in many cases, efficiency and termination
properties.  
Let us consider a program to compute distances between nodes in a
graph written using tabling (Fig.~\ref{fig:dist}, left). The query
\code{?-dist(a,Y,D), D<K.} would loop under SLD due to the
left-recursive rule, while it would terminate under tabling for
acyclic graphs.

Tabling records the first occurrence of each call to a tabled
predicate (the \emph{generator}) and its answers.  In variant tabling
(the most usual form of tabling), when a call is found to be equal,
modulo variable renaming, to a previous generator, the execution of the
call is suspended and it is flagged as a \emph{consumer} of the
generator.  For example \code{dist(a,Y,D)} is a variant of
\code{dist(a,Z,D)} if \code{Y} and \code{Z} are free variables.  Upon
suspension, execution switches to evaluating another untried branch.
A branch which does not suspend
can generate  answers for the initial goal.
When a generator finitely finishes exploring all the clauses and all
answers are collected,  the consumers that depend on it are resumed
and fed with the 
answers of the generator. This may make generators produce new answers
which can in turn resume more consumers.  This process finishes when
no new answers can be generated --- i.e., a fixpoint has been reached.
Tabling is sound and, for programs with a finite Herbrand model, 
complete (and, therefore, it always finishes in these cases).

\begin{figure}
  \begin{subfigure}[b]{.28\linewidth}
\begin{lstlisting}[style=MyProlog]
:- @table dist/3@.

dist(X,Y,D) :- 
    dist(X,Z,D1), 
    edge(Z,Y,D2), 
    D is D1+D2.
dist(X,Y,D) :-
    edge(X,Y,D).
\end{lstlisting}
    \caption{Tabling}
  \end{subfigure}
  \begin{subfigure}[b]{.3\linewidth}
\begin{lstlisting}[style=MyProlog]
:- @table dist/3@.

dist(X,Y,D) :- 
    D1.>.0, D2.>.0, 
    D.=.D1+D2, 
    dist(X,Z,D1),
    edge(Z,Y,D2).
dist(X,Y,D) :- 
    edge(X,Y,D).
\end{lstlisting}
    \caption{TCLP}
  \end{subfigure}
  \begin{subfigure}[b]{.35\linewidth}
\begin{lstlisting}[style=MyProlog]
:- @table dist(_,_,min)@.

dist(X,Y,D) :- 
    dist(X,Z,D1), 
    edge(Z,Y,D2), 
    D is D1+D2.
dist(X,Y,D) :-
    edge(X,Y,D).
\end{lstlisting}
    \caption{\atclp}
  \end{subfigure}
  \caption{Distance traversal in a graph. Note: The symbols \texttt{\#>} and
    \texttt{\#=} are (in)equalities in CLP.}
  \label{fig:dist}
\end{figure}

However, in a cyclic graph, \code{dist/3} has an infinite Herbrand
model: every cycle can be traversed repeatedly and create paths of
increasing length. Therefore, the previous query %
\mbox{\code{?-dist(a,Y,D), D<K}} will not terminate under variant
tabling, although the query as a whole has a finite model.

On the other hand, if the integration of tabling and CLP
(Fig.~\ref{fig:dist}, center) uses \emph{constraint
  entailment}~\cite{chico-tclp-flops2012-short}, calls to
\code{dist/3} will suspend if there are previous similar calls that
are more general, and only the most general answers will be kept.  The query
\mbox{\code{?-D.<.K, dist(a,Y,D)}} 
terminates under TCLP because by placing the constraint \code{D.<.K}
before \code{dist(a,Y,D)}, the search is pruned 
when the values in \code{D} are larger than or equal to
\code{K}.

This illustrates the main idea underlying the use of entailment
($\sqsubseteq$) in TCLP: more particular calls (consumers) can suspend
and later reuse the answers collected by more general calls
(generators).
In order to make this entailment relationship explicit, we will
represent a TCLP goal as \mbox{\goaltt{g}{c_{g}}} where $g$ is the
call (a literal) and $c_g$ is the projection of the current constraint
store onto the variables of the call.
For example, \goaltt{dist(a,Y,D)}{D>0 \land D<75} 
entails the goal %
\goaltt{dist(a,Y,D)}{D<150} %
because \mt{(D>0 \land D<75) \sqsubseteq D<150}.
The latter is therefore more general (i.e., it is a generator) than
the former (a consumer).
%
All the solutions of a consumer are solutions for its generator, since
the space of solutions of the consumer is a subset of that of the
generator.  However, not all answers from a generator are valid for
its consumers.  For example \mt{Y=b \land D>125 \land D<135} is a
solution for our generator, but not for our consumer, since the
consumer call was made under a constraint store more restrictive than
the generator.  Therefore, the tabling engine has to filter, via the
constraint solver, the answers from the generator that are consistent
w.r.t.\ the constraint store of the consumer.

Additionally, the \mtclp framework~\cite{TCLP-tplp2019} has been used
to implement in Ciao a framework, called \atclp~\cite{atclp-padl2019},
that incrementally computes aggregates for elements in a lattice. The
\atclp framework uses the entailment and join relations in a lattice to
define and compute aggregates, and to decide whether some atom is
compatible with (i.e., entails) the aggregate.
For example, the directive \code{:-table dist(_,_,min)} 
(Fig.~\ref{fig:dist}, right), specifies the (aggregate) mode \code{min}
for the third argument.  The query \code{?-dist(a,Y,D)} will in this
case terminate because only the shortest distance between two nodes
found at every moment is kept, and it will be returned in \code{D} as
a result of the evaluation of the initial call.
Other tabling engines implement \emph{answer
  subsumption}~\cite{swift2010:subsumption} or a restricted form of it
via \emph{mode-directed
  tabling}~\cite{tabling-modes-GuoG08,zhou2010mode,swi-journal-2012,yap-journal-2012}, that can be used
to compute aggregates.  However, answer subsumption, as implemented in
XSB, assumes answers to be safe (i.e., ground) and works on non-ground
answers only in some cases, so it would in principle not be applicable
when answers are constraints.  Answer subsumption also performs
subsumption only on answers, while Aggregate-TCLP can in addition
check entailment for calls.  In the case of the TCLP implementation of
the abstract interpreter, this  makes it possible to reuse answers
obtained from calls semantically equivalent (i.e., calls whose
associated abstract substitutions differ, but that still represent the
same object in the lattice) and/or more general (i.e., that represent
an element higher in the lattice hierarchy).  Note that in our
benchmarks we are using semantic equivalence, since using entailment
to detect more general calls would cause a loss of precision as the
domains we are using are non-relational. Last, answer subsumption does
not provide the freedom to be used with aggregates that cannot be
expressed in terms of a lattice, such as \code{sum/3},
which~\cite{atclp-padl2019} can work around.


\subsection{The PLAI algorithm}
\label{sec:plai-fram-ciaopp}

We assume that the reader is familiar with the basic principles of
abstract interpretation~\cite{Cousot77,bruy91,prog-analysis-book}.
The PLAI algorithm used by the abstract interpreter of CiaoPP for
static analysis extends the fixpoint algorithms proposed
by~\cite{bruy91} with the optimizations described
in~\cite{mcctr-fixpt}.  In logic programming, all possible concrete
substitutions in the program (i.e., terms to which the variables in
that program will be bound at run-time for a given query) can be
infinite, which gives rise to an infinite execution tree.
The core idea of PLAI is to represent this infinite execution tree by an
abstract and-or tree using abstract substitutions to finitely represent
the possibly infinite sets of substitutions in the concrete domain. The
set of all possible abstract substitutions that a variable can be
bound to is the \emph{abstract domain} which is usually a complete
lattice (or a complete partial order of finite height).

\paragraph{\textbf{Domains in PLAI}}

PLAI is domain-independent: new abstract domains can be easily
implemented and integrated by using a common interface. The
operations required by the domain interface are:

\begin{itemize}
  \renewcommand{\u}{\vec{u}}
  \renewcommand{\l}{\lambda}
  \renewcommand{\b}{\beta}
\item \code{$\l '\ \sqcup\ \l ''$}, which gives the LUB of the abstract
  substitutions $\l '$ and $\l ''$. The LUB operation is defined in
  terms of the $\sqsubseteq$ relation of the abstract domain.
\item \code{call_to_entry(p($\u$),C,$\l$)}, where \code{C} is a clause
  and \code{p($\u$)} is a call.  It gives an abstract
  substitution describing the effects on \code{vars(C)} of unifying
  \code{p($\u$)} with \code{head(C)} given an abstract substitution $\l$
  for the variables in $\u$.
\item \code{exit_to_success($\l$, p($\u$), C, $\b$)} which returns an
  abstract substitution describing the effect of execution
  \code{p($\u$)} against clause \code{C}. For this, the variables of
  the abstract substitution $\b$ are renamed taking into account the
  unification with the terms in \code{head(C)} and the variables in
  \code{p($\u$)}, and a new abstract substitution is returned updating
  $\l$ with the new information.
\item \code{extend($\l$,$\l'$)} which extends abstract substitution $\l$
  to incorporate the information in $\l$' in a way that it is still
  consistent.
\item \code{project_in($\u$,$\l$)} which extends the abstract substitution $\l$
  so that it refers to all the variables in $\u$.
\item \code{project_out($\u$,$\l$)} which restricts the abstract
  substitution $\l$ to refer only to the variables in $\u$.
\end{itemize}

For additional examples of abstract domains integrated in CiaoPP, we refer
the reader
to~\cite{nfplai-flops04-short,abs-int-naclp89,eterms-sas02}.

\paragraph{\textbf{And-Or trees and substitutions}}

In PLAI, the abstract and-or tree is constructed using a top-down
driven strategy (instead of a bottom-up computation) so that the
computation is restricted to what is required for the given query.
In the resulting and-or tree, an \emph{and-node} is a clause head
\code{h} whose children are the literals in its body,
\code{p$_1$,$\dots$,p$_n$}, and
an \emph{or-node} is a literal, \code{p$_i$}, whose children are
the heads \code{h$_1$,$\dots$,h$_m$} of the clauses that unify with
\code{p$_i$}. 
Its construction starts with the abstract call substitution for the
query. Then, abstract substitutions at all points of the abstract
and-or tree are computed and finally, the success substitution for the
query is computed.

Inside a clause, abstract substitutions at every point are denoted
depending on their position among its literals. Given a clause
\code{h:-p$_1$,$\dots$,p$_n$}, let $\l_i$ and $\l_{i+1}$ be the
abstract substitutions to the left and right of the subgoal $p_i$,
$1 \leq i \leq n$.
Then, $\l_i$ and $\l_{i+1}$ are, respectively, the \emph{abstract call
  substitution} and the \emph{abstract success substitution} for the
subgoal $p_i$.
The projection of $\l_1$ on $vars(\mathtt{h})$ is the \emph{abstract
  entry substitution}, $\b_{entry}$, of the given clause, and, similarly, the
projection of $\l_{n+1}$ on $vars(\mathtt{h})$ is its \emph{abstract exit substitution}, $\b_{exit}$.
%
The abstract substitutions for a clause are computed as follows:

\begin{algorithm}[t]
  \renewcommand{\u}{\vec{u}}
  \renewcommand{\l}{\lambda}
  \renewcommand{\b}{\beta}
  \caption{\code{entry_to_exit}: Compute exit substitution from entry
    substitution.}
  \label{alg:entry}
  \KwData{A clause C of the form \code|h($\u$):-p$_1$($\u_1$),$\dots$,p$_m$($\u_m$)|;
      an entry substitution $\b_{entry}$}
    \KwResult{An exit substitution $\b_{exit}$}
    $\l_1$ := \code{project_in($vars(C)$,$\b_{entry}$)}\;
    \For{i := 1 to m}{
      $\l_{i+1}$ := \code|call_to_success(p$_i$($\u_i$), $\l_i$)|\;
    }
    \Return{\code|project_out($\u$, $\l_{m+1}$)|\;}
\end{algorithm}

\begin{algorithm}[t]
  \renewcommand{\u}{\vec{u}}
  \caption{\code{call_to_success}: Compute success substitution
    from call substitution.}
  \label{alg:call}
 \KwData{A goal \code|p($\u$)|; an abstract call substitution $\lambda_{call}$}

  \KwResult{A success substitution $\lambda_{success}$}
   $\lambda_{proj}$ := \code|project_out($\u$,$\lambda_{call}$)|\;
   $\lambda'$ := $\bot$\;
  \For{each clause C which unifies with p($\u$)}{
   $\beta_{exit}$ := \code|entry_to_exit(C, call_to_entry(p($\u$), C, $\lambda_{proj}$))|\;
  $\lambda'$ := $\lambda'$ $\sqcup$
  \code|exit_to_success($\lambda_{proj}$, p($\u$), C, $\beta_{exit}$)|\;
  }
 \Return{\code|extend($\lambda_{call}$, $\lambda$|$'$\code|)|\;}
\end{algorithm}

\begin{itemize}
\item Exit substitution from the entry substitution
  (Algorithm~\ref{alg:entry}):
  Given a clause \code{h:-p$_1$,$\dots$,p$_n$} and an entry
  substitution $\b_{entry}$ for the clause head $h$, the call
  substitution $\l_1$ for $p_1$ is computed by simply adding to
  $\b_{entry}$ an abstraction for the variables in the clause that do
  not appear in the head. The success substitution for $p_1$ is
  $\l_2$, and it is computed as explained below (essentially, by
  repeating this same process for the clauses which unify with $p_1$).
  $\l_3, \dots, \l_{n+1}$ are computed similarly. The exit
  substitution $\b_{exit}$ for this clause is the projection of
  $\l_{n+1}$ onto $\vec{u}$, the variables in $h$.
\item Success substitution from the call substitution
  (Algorithm~\ref{alg:call}):
  Given a call substitution $\l_{call}$ for a subgoal $p$, let
  $h_1, \dots, h_m$ be the heads of clauses that unify with
  $p$. Compute the entry substitutions
  $\b1_{entry}, \dots, \b m_{entry}$ for these clauses.  Compute their
  exit substitutions $\b1_{exit}, \dots, \b m_{exit}$ as explained
  above. Compute the success substitutions
  $\l1_{success}, \dots, \l m_{success}$ from the exit substitutions
  corresponding to these clauses. At this point, all different success
  substitutions can be considered for the rest of the analysis, or a
  single success substitution $\l_{success}$ for subgoal $p$ computed
  by means of an aggregation operation for
  $\l1_{success}, \dots, \l m_{success}$. This aggregate is the least
  upper bound (LUB), denoted by $\sqcup$, of the abstract domain.
\end{itemize}

Note that these two procedures are mutually recursive and would not
finish in case of mutually recursive calls.  They merely describe how
abstract substitutions are generated for the case of literals in a
body (by carrying success abstract substitutions to call abstract
substitutions) and how entry and exit substitutions of several clauses
are composed together.  For the general case of recursive predicates,
where repeated calls and termination have to be detected, PLAI
implements a fixpoint algorithm that we sketch below.

\paragraph{\textbf{PLAI's fix point algorithm}}

The core idea of PLAI's fixpoint algorithm~\cite{mcctr-fixpt} is that
the subtree corresponding to the abstract interpretation of a node
with a recursive predicate \code{p} should be finite.
If the abstract domain is finite, a predicate \code{p} can only have a
finite number of distinct call substitutions and therefore the subtree
can only have a finite number of occurrences of nodes that have a
variant of \code{p} and which themselves have subtrees.
In addition to that, all other nodes in the subtree with the same
predicate name \code{p} and with the same call substitutions (modulo
variable renaming) use the approximate value of the success
substitution computed previously for the root node of the subtree
labeled with \code{p}, and hence they do not have any descendent
nodes.

Based on this idea, the fixpoint algorithm iteratively refines the approximate
values of the success substitution of the recursive predicate \code{p}
as follows:

\begin{itemize}
\item First, it computes an approximate value of the projected success
  substitution using the LUB of the projected success substitutions
  corresponding to the \textbf{non-recursive} clauses of \code{p}.
  This provides an initial, hopefully non-empty, abstract substitution
  that is fast to compute (it does not need to check for repeated
  calls or termination) and accelerates the convergence of the
  fixpoint algorithm.  In practice, it can be delegated to a
  specialized version of Algorithms~\ref{alg:entry} and~\ref{alg:call}
  restricted to non-recursive calls / clauses.  These can be
  determined beforehand by a reachability analysis based on 
  strongly connected components.

\item Then, it traverses the (finite) subtree corresponding to
  \code{p}
  in a depth-first fashion.  When an entry-exit combination is needed
  for a call to \code{p} having the same call substitution (modulo
  variable renaming), the existing approximation is used.
  For a call to \code{p} with a different call substitution, a new
  (nested) fixpoint computation is started.  When the analysis returns
  to the root of the subtree, the success substitution for \code{p} is
  updated as the LUB of the previous value and the value just computed
  from the recursive clauses of \code{p}.
\item If there is a change in the success substitution for \code{p},
  the depth-first traversal is restarted using the new success
  substitution, which is used for the subtree nodes corresponding to
  \code{p} that have a compatible call substitution.
  These depth-first traversal iterations can take place only a bounded
  number of times, since the LUB operation is monotonic and the
  abstract substitutions form a lattice of finite
  height.\footnote{While it is true that abstract domains can be
    infinite, if convergence is not reached after some time, a
    widening operation changes the representation of the abstract
    substitutions to a coarser domain that has more chances to
    converge (or is sure to converge, if it is finite).}  Therefore, a
  fixpoint will be reached in a finite number of steps.
\item If there is no change in the success substitution for the root
  node of the subtree of \code{p} for a given call substitution, then
  the analysis of that subtree is complete (for that call
  substitution) and the fixpoint computation of the predicate \code{p}
  terminates. 
\end{itemize}

For recursive predicates called from within recursive predicates, the
dependencies between nested calls have to be recorded to restart the
traversal of the subtrees containing  predicate calls whose success
substitution has been updated.

\section{Implementations of the PLAI Algorithm: Prolog vs.\ Tabling}

We will now describe more in depth how the PLAI algorithm is
implemented in CiaoPP\footnote{The code is available at
  \url{www.ciao-lang.org}.  For the reader convenience, we sketch it
  in~\ref{sec:Ap-plai-CiaoPP}.} and highlight the differences w.r.t.\
the version that uses Tabled CLP.

\subsection{PLAI in CiaoPP}
\label{sec:plai-ciaopp}

The implementation of \code{call_to_success} is the entry point, as it
relates the entry and exit substitutions of a call (in
particular, of the top-level call).
%
%
%
During the analysis of a goal \code|p($\u$)|, and for each clause that
unifies with \code|p($\u$)|, the predicate \code{call_to_success} invokes
\code{entry_to_exit} which, for each subgoal in the body of the
clause, invokes again \code{call_to_success}. 
The abstract interpreter is able to stop the evaluation of a part of
the program and move to another part to evaluate calls to other
predicates.
%
%
The implementation of PLAI is optimized to accelerate the convergence
of the fixpoint and reduce the computation by reusing previous
results, among other techniques.

The PLAI algorithm is based on the construction of an and-or tree,
described in Section~\ref{sec:plai-fram-ciaopp}, with the nodes
representing the predicate calls
visited during the analysis. To construct this tree,
\code{call_to_success} identifies each goal with its corresponding
and/or node and with the specialized version of its father
(i.e., the version of the literal that originated the call) and
carries around a list with the nodes on which the current goal depends.
The analysis starts with a query (a goal) and a call
substitution. With this information, \code{call_to_success} creates
the root node of the tree and the list of
clauses that unify with the goal.
If the goal corresponds to a non-recursive predicate, it computes the
success substitution which is asserted in a memo-table to reuse the
result later on.
Otherwise, the goal corresponds to a recursive predicate and it is
dealt with by the fixpoint algorithm: first, it evaluates the
non-recursive clauses obtaining an approximation of the success
substitution and, after this, it starts the fixpoint computation.

During the fixpoint computation, for a goal with a given call
substitution:

\begin{itemize}
\item If complete information has been already inferred and saved,
  \code{call_to_success} reuses it, to avoid re-computations.
\item If it is already inside a fixpoint computation (some parent
  started a fixpoint with the same call), \code{call_to_success}
  reuses the approximation stored for this call, to avoid entering
  loops.
\item If an analyzed call depends on other nodes whose fixpoint
  are not completed yet, 
  two cases are treated:
  \begin{itemize}
  \item If the information on which the predicate depends is updated,
    a local fixpoint computation is started.
  \item Otherwise, nothing is done.
  \end{itemize}
  To decide whether updated information for a node is available, the
  information inferred for it has a version number:
  \begin{itemize}
  \item When the information on a node is updated, its version number
    is increased by one.
  \item When a node uses information from another node, it stores the
    version of that information in the list of nodes on which it depends.
  \end{itemize}
  Version numbers are used to detect updates of the information on
  which a node analysis depend.  If the version number of the last
  information used from a node does not match its current version
  number, there has been an update that needs to be propagated.

\end{itemize}

When the fixpoint computation finishes and the list of dependent nodes
is empty, the current information for this call is asserted.
Otherwise, if this list is not empty, the information remains flagged
as an approximation and the fixpoint restarts.  As it can easily be
seen, while the algorithm can be conceptually not too complex, its
implementation is cumbersome and at points costly, since many
interactions are done through the database using identifiers for
program points.

\subsection{The PLAI Algorithm in TCLP}
\label{sec:call-enta-check}



The PLAI code using tabling is a simplification of the corresponding
Prolog implementation.  The main points that were changed are:

\begin{itemize}
\item The handling of dependencies among nodes and the detection of
  termination in the fixpoint computation, that were explicit in the
  Prolog version, are now transferred to the underlying fixpoint of
  the tabling engine.
\item The calculation of the LUB of the abstract substitutions
  generated by different clauses unifying with a call is done via
  lattice-based constraint aggregation (which is in turn built upon
  tabling).
\end{itemize}


\subsubsection{Internal Database and Dependencies}

In the Prolog implementation, the information related to the abstract
substitutions is kept in a dynamic database relating code, program
points, entry/exit substitutions, and dependencies.  This makes it
globally accessible and allows it to survive across backtracking and calls, so that
it does not need to be carried around the program and be rebuilt every
time there is a change in the substitution at a program point.
%

However, making the abstract interpreter update that information,
switch among calls, and re-analyze calls needs accessing and updating
this database, which is costly and mixes declarative and imperative
styles.
On top of that, the CiaoPP implementation has been fine-tuned during
many years to avoid unnecessary (re-)analyses and minimize
the overhead of accessing the database.  All of these optimizations
cause the code to have to deal with specific cases for the sake of
performance, hence adding to its complexity.
But despite the involved implementation, this machinery mimics, at
Prolog level, an infrastructure similar to a tabling engine, but
specialized for a given program ---the abstract interpreter--- and
with  optimizations specific for the task at hand.

This bookkeeping becomes unnecessary when using a tabling-based
implementation.  An abstract interpreter written using tabling and
equipped with the capability to detect when two syntactically
different substitutions represent the same object, can automatically
take care of termination, suspend analysis when repeated calls are
detected, and resume them when new information is available --- all of
it as part of the normal execution of a tabled program, without having
to explicitly update and check dependencies.


That makes the code much simpler (no dependencies, lists of pending
goals, resuming, etc.\ need to be explicitly coded) and shorter (we
have obtained a threefold reduction in code size).  On the other hand,
the tabling engine is generic and cannot decide which suspension
and/or resumption policy is better for a particular application.  We
on purpose chose to (a) keep the TCLP code simple and not include any
specific heuristic in the code, (b) not to reimplement an analyzer
from scratch, but simplify existing code, and (c) keep exactly the
same interfaces (both those offered to the rest of CiaoPP and those
required by the fixpoint code) so that the TCLP-based abstract
interpreter can interoperate with the rest of the CiaoPP machinery as
a drop--in replacement with close to zero effort.  For these and other
reasons, our performance figures (Section~\ref{sec:evaluation}) are a
lower bound of what could be achieved.

\begin{figure}
\begin{lstlisting}[style=MyProlog]
call_to_success(SgKey,Call,Proj,Sg,Sv,AbsInt,Succ) :-  
	call_to_success_fixpoint(SgKey,Sg, st(Sv,Call,Proj,AbsInt,Prime) ),
	each_extend(Sg,Prime,AbsInt,Sv,Call,Succ).

:- @use_package(tclp_aggregate)@.
:- @table call_to_success_fixpoint(_,_,abst_lub)@.
call_to_success_fixpoint(SgKey,Sg, st(Sv,Call,Proj,AbsInt,Prime) ) :-
	trans_clause(SgKey,_,Clause),
	do_nr_cl(Clause,Sg,Sv,Call,Proj,AbsInt,Prime).
call_to_success_fixpoint(SgKey,Sg, st(Sv,_Call,Proj,AbsInt,Prime) ) :-
	\+ trans_clause(SgKey,_,_),
	apply_trusted0(Proj,SgKey,Sg,Sv,AbsInt,_ClId,Prime).
\end{lstlisting}
  \caption{Implementation of \code{call_to_success/7} under the
    \gtclp framework}
  \label{fig:call_ent}
\end{figure}

As an example, the implementation of \code{call_to_success/13} in
Prolog checks several cases: if the call being analyzed is complete,
under evaluation in a fixpoint,  a call to a recursive predicate,
a call to a non-recursive predicate, etc.\ to update information
accordingly.  It eventually invokes \code{proj_to_prime_nr/9}, which
starts the fixpoint computation itself, and which recursively calls
\code{call_to_success/13}.  \code{call_to_success/13} has eight
clauses and \code{proj_to_prime_nr/9} has six clauses
(see~\ref{sec:Ap-plai-CiaoPP} or the corresponding file at
\papermaterial).

In the tabling implementation, the underlying engine and the calls to
the abstract domain operations through the constraint solver interface
take care of these cases and dependencies.  This makes the
implementation of \code{call_to_success} have just \textbf{one} clause
(Fig.~\ref{fig:call_ent}).  The counterpart to
\code{proj_to_prime_nr/9} (which we renamed
\code{call_to_success_fixpoint/3} for clarity) has just two clauses:
one for user predicates and another one for library and builtin
predicates.

%


Additionally, the use of tabling makes it unnecessary to save
explicitly all the intermediate substitutions, database identifiers
for calls and program points, dependencies among goals, etc.  This
reduces the number of arguments, and \code{call_to_success}
went from thirteen used in Prolog:
\begin{lstlisting}[style=MyProlog, numbers=none]
call_to_success(RFlag,SgKey,Call,Proj,Sg,Sv,AbsInt,ClId,Succ,List,F,N,Id)
\end{lstlisting}

\noindent
to seven in the tabling-based implementation:
\begin{lstlisting}[style=MyProlog, numbers=none]
call_to_success(SgKey,Call,Proj,Sg,Sv,AbsInt,Succ)
\end{lstlisting}


\subsubsection{Deciding Termination and Computing the LUB}

In the PLAI algorithm, the different exit substitutions obtained from
the clauses that unify with a given call are combined using the LUB
operator of the abstract domain (Algorithm~\ref{alg:call}): exit
substitutions $\b_{i\;exit}$, for every clause $C_i$ are joined to
return the success substitution $\l_{success}$.

The  CiaoPP  implementation uses
\code{bagof/3}  to collect
all the clauses in a list and then traverses it and analyzes every
clause to create another list of abstract substitutions that are
joined with the LUB.  This processing is conceptually simple, but its
implementation obscures the code with low-level operations, does not
match the idea of having an interpreter executing on an abstract
domain, and requires database accesses to retrieve the substitution
applicable at that point.

In our implementation, the use of lattice-based aggregates with the
tabling engine~\cite{atclp-padl2019} simplifies the code.  The
\code{abst_lub} identifier in line 6 of Fig.~\ref{fig:call_ent} is the
name of an interface that has several missions: determine suspension of
calls, detect termination of the fixpoint, and perform aggregation of
abstract substitutions.
In the same line, the underscores state that the corresponding
arguments are to be checked for equality (necessary to decide whether
a fixpoint has been reached) using the \emph{variant} policy, i.e.,
syntactical equality modulo variable renaming.

The implementation of the interface named \code{abst_lub} in
Fig.~\ref{fig:ent_interface} tells the tabling engine how to treat the
argument selected previously with this identifier.  In particular, the
tabling engine checks the corresponding arguments for equality by
calling \code{call_entail/3}.  In our case, two abstract substitutions
are termed equal if the abstract domain implementation
(\code{identical_abstract/3}) decides so.  This makes it possible to
detect that two different representations correspond to the same
object in the lattice and, if so, suspend a call or retrieve saved
answers for it.

\begin{figure}
\begin{lstlisting}[style=MyProlog]
call_entail(abst_lub, st(Sv,_,ProjA,AbsInt,_), st(Sv,_,ProjB,AbsInt,_) ) :-
	identical_abstract(AbsInt,ProjA,ProjB).
answer_entail(abst_lub, st(Sv,_,_,AbsInt,PrimeA), st(Sv,_,_,AbsInt,PrimeB) ) :-
 	less_or_equal(AbsInt,PrimeA,PrimeB).
answer_join(abst_lub,st(Sv,_,_,Abs, A), st(Sv,_,_,Abs, B), st(Sv,_,_,Abs,New) ) :-
        compute_lub(Abs,[A,B],New).
apply_answer(abst_lub, st(Sv,_,_,AbsInt,Prime), st(Sv,_,_,AbsInt,Prime) ).
\end{lstlisting}
  \caption{Code of the operator \code{abst_lub} under the \gtclp
    framework}
  \label{fig:ent_interface}
\end{figure}


The code in Fig.~\ref{fig:ent_interface} also aggregates the results
returned in the third argument (the abstract substitutions) by joining
them with the LUB of the lattice.  The tabling engine calls
\code{answer_entail/3} to decide whether a new answer (a substitution)
is or not more general than an existing answer
(\code{less_or_equal/3}).  If its not comparable,
\code{answer_join/4} (which in turn invokes \code{compute_lub/3}) is
called to compute the LUB of a previous answer and the new one.
%
%
%
With these definitions, lines 7 to 12 in Fig.~\ref{fig:call_ent}
contain \textbf{all} the code necessary to return the exit
substitution of a call w.r.t.\ all its matching clauses.  The
implementation of the LUB operation (\code{abs_lub},
Fig.~\ref{fig:ent_interface}) is based on the operations
provided by the abstract domain  implementation.

This code also performs an incremental computation of the LUB as follows:
%
%
upon success, the first answer, corresponding to the exit substitution
$\b1_{exit}$, is stored in the answer table of the tabled predicate.
Let us call this stored answer $\b_{exit}$.
For the subsequent exit substitutions \mbox{$\b\,i_{exit}, i > 1$}, there are
two possible cases: if the saved substitution is more general
($\b\,i_{exit} \sqsubseteq \b_{exit}$), then $\b\,i_{exit}$ is
discarded; otherwise we make $\b_{exit} = \b_{exit} \sqcup \b\,i_{exit}$.
%



%

\subsubsection{Connecting Abstract Substitutions with Lattice-Based
  Aggregates} 

The TCLP system handles entailment, aggregation, etc.\ by delegating
operations to an underlying constraint solver using a fixed
interface~\cite{TCLP-tplp2019}.
%
Since we purposely did not change the representation of the CiaoPP
abstract domains (they are used in other parts of the system), we
constructed a bridge between these domains and the interface that TCLP
expects.

The original entry point of the fixpoint, \code{proj_to_prime_nr/9}
(renamed as \code{call_to_success_fixpoint/3} in the TCLP
implementation), now tabled, is automatically rewritten (by the
package \code{tclp_aggregate}) to call an auxiliary predicate that, at
run time, substitutes the arguments carrying abstract substitutions by
attributed variables~\cite{holzbaur-plilp92} that simulate having a
constrained variable.  Their attributes are tuples that contain (a)
the identifier (\code{abst_lub}, in our example) that determines the
interface to be used and (b) the abstract substitution and ancillary
information necessary by the abstract interpreter.

When one operation of the tabling engine involves a call with
attributed variables, the engine checks if it has an attribute with
contents it recognizes.  If so, it calls the corresponding predicate
from the interface that, in our case, operates on the substitution
stored in the attributes.

\section{Evaluation}
\label{sec:evaluation}

Besides simplifying code, the implementation of PLAI using TCLP gives
performance advantages in many cases.  These come mainly because part
of the bookkeeping related to dependencies, saving the analysis state
when restarting the analysis of a dependent call, checking for
termination, etc.\ are handled at a lower level.  On the other
hand, the implementation currently in CiaoPP, as commented before, has
been fine-tuned and specialized during many years to minimize the
overhead of the fixpoint implementation, so that a large proportion of
the analysis
time is spent in domain-related operations.  On top of that, the
CiaoPP domain representation and domain operations are designed to
work well with its current architecture and coding decisions
(e.g. saving and retrieving from the dynamic databases) and are
suboptimal for a tabling-based implementation: for example, redundant
data is manipulated and/or stored.  As commented earlier, we did not
change any of these so the TCLP fixpoint can seamlessly interact with
the rest of the CiaoPP tool, exposing and using exactly the same
interfaces.

\begin{table}[t]
  \begin{tabular}{lrrr}
    \toprule
    & Speedup & TCLP (ms)  & CiaoPP (ms) \\
    \midrule
    fibf\_alt    & 1.60 & \textbf{0.29} & 0.46          \\
    aiakl        & 1.56 & \textbf{2.45} & 3.82          \\
    boyer        & 1.50 & \textbf{7.31} & 10.97         \\
    pv\_queen    & 1.46 & \textbf{0.74} & 1.07          \\
    subst        & 1.41 & \textbf{0.25} & 0.35          \\
    pv\_gabriel  & 1.37 & \textbf{3.65} & 4.99          \\
    rdtok        & 1.32 & \textbf{7.03} & 9.25          \\
    mmatf        & 1.24 & \textbf{0.31} & 0.39          \\
    hanoi        & 1.22 & \textbf{0.53} & 0.65          \\
    revf\_lin    & 1.20 & \textbf{0.27} & 0.32          \\
    append       & 1.20 & \textbf{0.17} & 0.20          \\
    rev\_lin     & 1.19 & \textbf{0.26} & 0.31          \\
    prefix       & 1.16 & \textbf{0.27} & 0.31          \\
    revf         & 1.15 & \textbf{0.32} & 0.37          \\
    pv\_plan     & 1.15 & \textbf{1.94} & 2.23          \\
    sublist\_app & 1.14 & \textbf{0.24} & 0.27          \\
    reverse      & 1.14 & \textbf{0.38} & 0.43          \\
    flatten      & 1.13 & \textbf{0.55} & 0.62          \\
    palindro     & 1.12 & \textbf{0.34} & 0.38          \\
    fact         & 1.08 & \textbf{0.25} & 0.27          \\
    rotate       & 1.06 & \textbf{0.46} & 0.49          \\
    maxtree      & 0.98 & 0.63          & \textbf{0.61} \\
    zebra        & 0.92 & 1.38          & \textbf{1.26} \\
    browse       & 0.89 & 1.76          & \textbf{1.57} \\
                 &      &               &                \\
    AVG          & 1.31 & 31.78         & 41.59          \\
    \bottomrule
  \end{tabular}
 \caption{Performance comparison: CiaoPP fixpoint in Prolog and TCLP
   (\emph{Groundness} domain).}
 \label{tab:gr}
\end{table}

\begin{table}[t]
  \begin{tabular}{lrrr}
    \toprule
    & Speedup & TCLP (ms) & CiaoPP (ms) \\
    \midrule
    fact         & 1.30 & \textbf{0.26} & 0.33           \\
    pv\_queen    & 1.23 & \textbf{1.21} & 1.49           \\
    mmatf        & 1.17 & \textbf{0.51} & 0.60           \\
    mmatrix      & 1.15 & \textbf{0.53} & 0.61           \\
    prefix       & 1.14 & \textbf{0.46} & 0.52           \\
    revf         & 1.12 & \textbf{0.47} & 0.53           \\
    revf\_lin    & 1.10 & \textbf{0.39} & 0.43           \\
    reverse      & 1.10 & \textbf{0.39} & 0.43           \\
    rev\_lin     & 1.10 & \textbf{0.38} & 0.42           \\
    rotate       & 1.06 & \textbf{0.72} & 0.76           \\
    pv\_pg       & 1.01 & \textbf{2.67} & 2.70           \\
    append       & 0.98 & 1.11          & \textbf{1.09}  \\
    sublist\_app & 0.96 & 0.87          & \textbf{0.84}  \\
    zebra        & 0.91 & 16.34         & \textbf{14.80} \\
                 &      &               &                \\
    AVG          & 0.97 & 26.31         & 25.55          \\
    \bottomrule
 \end{tabular}                                                                                   
 \caption{Performance comparison: CiaoPP fixpoint in Prolog and TCLP
   (\emph{Sh+Fr} domain).}
 \label{tab:shfr}
\end{table}

Even with these constraints, we observed speedups when analyzing most
programs from a benchmark set.
We used the \emph{Groundness} and \emph{Sharing+Freeness}~\cite{freeness-iclp91} domains due
to their relevance (e.g., for program optimization and correctness of
parallelization).
%
\emph{Groundness} (see Table~\ref{tab:gr} for performance results)
determines if some program variable will be bound to a ground term.
This is useful to derive modes, optimize unification, and improve the
precision of the \emph{Sharing+Freeness} analysis, among others.

\emph{Sharing+Freeness} (see Table~\ref{tab:shfr}) determines if two
(or more) program variables may be bound to terms sharing a common
variable.  It is useful to determine, for example, whether running two
goals in parallel may try to bind the same variable, thus causing
races and compromising correctness.
The benchmarks used are standard programs that have been  previously used to
evaluate CiaoPP.


%
All the experiments in this paper were performed on a Linux
5.0.0-13-generic machine with an Intel Core i7 at 1.80GHz with 16Gb of
memory and using \texttt{gcc 8.3.0} to compile the abstract machine of
Ciao Prolog.
In all cases, every program was analyzed 40 times and the 10 worst
times were discarded, both when using the tabling and the Prolog
implementation, to try to minimize the effect of spurious
interruptions, O.S.\ scheduling, etc.\ that can introduce noise in the
execution.  The remaining times were averaged.
All the code and the system under evaluation is available at
\papermaterial.

The average speedups in each table were calculated by adding up the
(averaged) execution times for all the benchmarks and dividing the
\emph{CiaoPP} time by the \emph{TCLP} time.  This shows that, on
average, the analysis with the \emph{Groundness} domain speeds up a
bit more than 30\%, while the analysis with the \emph{Sharing+Freeness}
has experienced, on average, a slight slowdown (about 3\%).

By looking at every benchmark in isolation, we can observe that the
speedups differ greatly among them.  We have sorted the benchmarks
according to the speedup to appreciate better the differences.  In
both cases, only a small part of the benchmarks (three) experienced a
slowdown, and even in these cases, the maximum slowdown was about
10\%.  In the case of \emph{Sharing+Freeness}, the slowest analysis
corresponded as well to the largest execution time (larger than the
rest of the benchmarks combined).  We want to note that this benchmark
(zebra) is probably not a representative of a typical program, as it
is a combinatorial problem with many free variables in
a single clause, some of which are aliased with each other.

The source of the speed difference is not easy to determine.  A
profile of the number of fixpoint calls in CiaoPP vs.\ fixpoint calls,
entailment checks, joins, etc. in the TCLP version does not seem to
show a correlation with the observed speedups.  We therefore
conjecture that the shape and size of the abstract substitution, and
the relative cost of checking entailment, has to be explored to have a
better explanation of the differences observed.

\section{Conclusions and Future Work}

We have presented a re-implementation of PLAI, a fixpoint computation
algorithm for abstract interpretation, using tabled constraint logic
programming.  The resulting code 
is considerably shorter than the current Prolog implementation of PLAI
in CiaoPP (one-third of its size) and much simpler: all the
bookkeeping necessary to keep track of dependencies between
predicates, analysis restarting, etc. is in charge of the tabling
engine, which increases the maintainability of the implementation of PLAI.


We have evaluated its performance using several benchmarks and
abstract domains, and compared it with the original implementation in
CiaoPP.  In most cases, the TCLP implementation showed improved
performance, sometimes with a speedup of 60\%.  In a few cases there
was a small slowdown, which we think is a reasonable price to pay for
the added code clarity, especially taking into account that there is
room for improvement in the current implementation.


Among the immediate future plans, we want to experiment
re-implementing the abstract domains with an optimized representation
of the abstract substitutions, and also use constraint logic
programming techniques to propagate the effects of updates.  We also
expect that, using constraints, we will be able to define widening
heuristics independently of the fixpoint algorithm thereby increasing
the resulting flexibility, precision and performance w.r.t.\ the state
of the art.

\subsection*{Acknowledgements}

We would like to thank Maximiliano Klemen, who helped us understand
the intricacies of the CiaoPP implementation of PLAI.  Thanks are also
due to Manuel Hermenegildo, who gave us very valuable feedback on the
paper manuscript and also a historical account on the early
relationship between tabling and efficient abstract interpretation
implementations.

\bibliographystyle{acmtrans}
\label{sec:bibliography}

\newpage
\appendix

\section{PLAI Algorithm Implementation Using TCLP}
\label{sec:Ap-plai-TCLP}

In this appendix we include the code
corresponding to the reimplementation of PLAI using TCLP.  It is not
expected to be used to understand the code (we did not add any
facility or improve its functionality), but rather to compare the code
length and complexity with that of the original PLAI in CiaoPP, which
we include in~\ref{sec:Ap-plai-CiaoPP}.  Therefore, we have
removed the comments that appear in the original files.  The files
with comments can be accessed at \papermaterial.

\vspace{1em}

\begin{lstlisting}[style=MyASP, mathescape = false, basicstyle =
\footnotesize\ttfamily\color{PrologPredicate}]
/*             Copyright (C)1990-2019 UPM-CLIP				*/

:- module(fixpo_plai_tabling,
	[
	    query/8,
	    init_fixpoint/0,
	    cleanup_fixpoint/1,
	    entry_to_exit/9
	],
	[assertions, datafacts]).

% Ciao library
:- use_module(engine(io_basic)).

:- use_module(library(aggregates), [bagof/3, (^)/2]).
:- use_module(library(lists), [member/2, append/3]).
:- use_module(library(terms_vars), [varset/2]).
:- use_module(library(terms_check)).
:- use_module(library(sets), [merge/3, ord_subtract/3]).
:- use_module(library(sort), [sort/2]).
:- use_module(library(messages)).
:- use_module(library(write)).

% CiaoPP library
:- use_module(ciaopp(preprocess_flags), [current_pp_flag/2, set_pp_flag/2]).

% Plai library
:- use_module(ciaopp(plai/fixpo_ops), [inexistent/2, variable/2, bottom/1, 
	singleton/2, fixpoint_id_reuse_prev/5, fixpoint_id/1, fixp_id/1,
	each_abs_sort/3,
	% each_concrete/4,
	each_extend/6, each_project/6, each_exit_to_prime/8, each_unknown_call/4,
	each_body_succ_builtin/12, body_succ_meta/7, reduce_equivalent/3,
	each_apply_trusted/7,	widen_succ/4,	decide_memo/6,clause_applies/2,
	abs_subset_/3]).

:- use_module(ciaopp(plai/domains)).
:- use_module(ciaopp(plai/trace_fixp), [fixpoint_trace/7, cleanup/0]).
:- use_module(ciaopp(plai/plai_db), 
	[ complete/7, memo_call/5, memo_table/6, cleanup_plai_db/1, patch_parents/6 ]).
:- use_module(ciaopp(plai/psets), [update_if_member_idlist/3]).
:- use_module(ciaopp(plai/re_analysis), [erase_previous_memo_tables_and_parents/4]).
:- use_module(ciaopp(plai/transform), [body_info0/4, trans_clause/3]).
:- use_module(ciaopp(plai/apply_assertions_old),
	[ apply_trusted0/7, 
 	  cleanup_trusts/1 ]).

:- doc(author,"Joaquin Arias").

:- doc(module,"This module adapts the implementation of the top-down
	fixpoint algorithm of PLAI using TCLP with aggregates and an
	extension that also checks call entailment.").

init_fixpoint.

cleanup_fixpoint(_AbsInt).

%------------------------------------------------------------------------%
% call_to_success(+,+,+,+,+,+,-)                                         %
%------------------------------------------------------------------------%

call_to_success(SgKey,Call,Proj,Sg,Sv,AbsInt,Succ) :-  
	call_to_success_fixpoint(SgKey,Sg, st(Sv,Call,Proj,AbsInt,Prime)),
	each_extend(Sg,Prime,AbsInt,Sv,Call,Succ).

%%%%%%%%%%%% TCLP interface %%%%%%%%%%%%%
 :- use_package(tclp_aggregate).
 :- table call_to_success_fixpoint(_,_,abst_lub).

call_entail(abst_lub, st(V,_,ProjA,AbsInt,_), st(V,_,ProjB,AbsInt,_)) :-
	identical_abstract(AbsInt,ProjA,ProjB),	!.

answer_entail(abst_lub, st(V,_,_,AbsInt,PrimeAs), st(V,_,_,AbsInt,PrimeBs),1) :-
	singleton(PrimeA,PrimeAs),
	singleton(PrimeB,PrimeBs),
 	less_or_equal(AbsInt,PrimeA,PrimeB), !.

answer_join(abst_lub,st(V,_,_,AbsInt,PrimeAs), st(V,_,_,AbsInt,PrimeBs),
	                                          st(V,_,_,AbsInt,PrimeNews)) :-
	singleton(PrimeA,PrimeAs),
	singleton(PrimeB,PrimeBs),
	singleton(PrimeNew,PrimeNews),
	compute_lub(AbsInt,[PrimeA,PrimeB],PrimeNew), !.

apply_answer(abst_lub, st(V,_,_,Ab,A), st(V,_,_,Ab,B)) :- A = B.

call_to_success_fixpoint(SgKey,Sg,st(Sv,Call,Proj,AbsInt,Primes)) :-
	trans_clause(SgKey,_,Clause),
	do_nr_cl(Clause,Sg,Sv,Call,Proj,AbsInt,Primes).
call_to_success_fixpoint(SgKey,Sg,st(Sv,_Call,Proj,AbsInt,Primes)) :-
	\+ trans_clause(SgKey,_,_),
	apply_trusted0(Proj,SgKey,Sg,Sv,AbsInt,_ClId,Prime),
	singleton(Prime,Primes).

do_nr_cl(Clause,Sg,Sv,Call,Proj,AbsInt,Primes):-
	Clause = clause(Head,Vars_u,K,Body),
	clause_applies(Head,Sg), !,
	varset(Head,Hv),
	sort(Vars_u,Vars),
	ord_subtract(Vars,Hv,Fv),
	process_body(Body,K,AbsInt,Sg,Hv,Fv,Vars_u,Head,Sv,Call,
	                                                Proj,Primes,_Id).
do_nr_cl(_Clause,_Sg,_Sv,_Call,_Proj,_AbsInt,[[]]).

process_body(Body,K,AbsInt,Sg,Hv,_Fv,_,Head,Sv,Call,Proj,LPrime,_Id):- 
	Body = g(_,[],'$built'(_,true,_),'true/0',true), !,
	singleton(Prime,LPrime),
	call_to_success_fact(AbsInt,Sg,Hv,Head,K,Sv,Call,Proj,Prime,_Succ).
process_body(Body,K,AbsInt,Sg,Hv,Fv,Vars_u,Head,Sv,_,Proj,Prime,Id):-
	call_to_entry(AbsInt,Sv,Sg,Hv,Head,K,Fv,Proj,Entry,ExtraInfo),
	singleton(Entry,LEntry),
	entry_to_exit(Body,K,LEntry,Exit,[],_,Vars_u,AbsInt,Id),
	each_exit_to_prime(Exit,AbsInt,Sg,Hv,Head,Sv,ExtraInfo,Prime).

%------------------------------------------------------------------------%
% entry_to_exit(+,+,+,-,+,-,+,+,+)                                       %
%------------------------------------------------------------------------%

entry_to_exit((Sg,Rest),K,Call,Exit,OldList,NewList,Vars_u,AbsInt,NewN):- !,
	body_succ(Call,Sg,Succ,OldList,IntList,Vars_u,AbsInt,K,NewN,_),
	entry_to_exit(Rest,K,Succ,Exit,IntList,NewList,Vars_u,AbsInt,NewN).
entry_to_exit(true,_,Call,Call,List,List,_,_,_):- !.
entry_to_exit(Sg,Key,Call,Exit,OldList,NewList,Vars_u,AbsInt,NewN):- 
	body_succ(Call,Sg,Exit,OldList,NewList,Vars_u,AbsInt,Key,NewN,_),
	true. 

body_succ(Call,_Atom,Succ,List,List,_HvFv_u,_AbsInt,_ClId,_ParentId,no):-
	bottom(Call), !,
	Succ = Call.
body_succ(Call,Atom,Succ,List,NewList,HvFv_u,AbsInt,ClId,ParentId,Id):- 
	Atom=g(Key,Sv,Info,SgKey,Sg),
	body_succ_(Info,SgKey,Sg,Sv,HvFv_u,Call,Succ,List,NewList,AbsInt,
	           ClId,Key,ParentId,Id).

body_succ_(Info,SgKey,Sg,Sv,HFv,Call,Succ,L,NewL,AbsInt,ClId,Key,PId,Id):-
	Info = [_|_], !,
	split_combined_domain(AbsInt,Call,Calls,Domains),
	map_body_succ(Info,SgKey,Sg,Sv,HFv,Calls,Succs,L,NewL,Domains,
	              ClId,Key,PId,Id),
	split_combined_domain(AbsInt,Succ,Succs,Domains).
body_succ_(Info,SgKey,Sg,Sv,HFv,Call,Succ,L,NewL,AbsInt,ClId,Key,PId,Id):-
	body_succ0(Info,SgKey,Sg,Sv,HFv,Call,Succ,L,NewL,AbsInt,
	           ClId,Key,PId,Id).

map_body_succ([],_SgKey,_Sg,_Sv,_HFv,[],[],L,L,[],_ClId,_Key,_PId,no).
map_body_succ([I|Info],SgKey,Sg,Sv,HFv,[Call|Calls],[Succ|Succs],L,NewL,
	      [AbsInt|Domains],ClId,Key,PId,Id):-
	body_succ0(I,SgKey,Sg,Sv,HFv,Call,Succ,L,_NewL,AbsInt,
	           ClId,Key,PId,_Id), !,
	map_body_succ(Info,SgKey,Sg,Sv,HFv,Calls,Succs,L,NewL,Domains,
	           ClId,Key,PId,Id).

body_succ0('$var',SgKey,Sg,_Sv_u,HvFv_u,Calls,Succs,List0,List,AbsInt,
        	_ClId,F,_N,_Id):-
        !,
	&( Calls=[Call],
	  concrete(AbsInt,Sg,Call,Concretes),
	  concretes_to_body(Concretes,SgKey,AbsInt,B)
	-> meta_call(B,HvFv_u,Calls,[],Succs,List0,List,AbsInt,_ClId,_Id,_Ids)
	 ; List=List0,
	   each_unknown_call(Calls,AbsInt,[Sg],Succs) % Sg is a variable
	&).
body_succ0('$meta'(T,B,_),SgKey,Sg,Sv_u,HvFv_u,Call,Succ,List0,List,AbsInt,
	    _ClId,_F,_N,_Id):-
        !,
	meta_call(B,HvFv_u,Call,[],Exits,List0,List,AbsInt,ClId,Id,_Ids),
	&( body_succ_meta(T,AbsInt,Sv_u,HvFv_u,Call,Exits,Succ) ->
	  true
	; % for the trusts, if any:
	  varset(Sg,Sv_r), % Sv_u contains extra vars (from meta-term)
                           % which will confuse apply_trusted
	  body_succ0(nr,SgKey,Sg,Sv_r,HvFv_u,Call,Succ,[],_List,AbsInt,
		        _ClId,_F,_N,_Id0)
	&).
body_succ0('$built'(T,Tg,Vs),SgKey,Sg,Sv_u,HvFv_u,Call,Succ,List0,List,AbsInt,
	    _ClId,_F,_N,_Id):-
	!,
	List=List0,
	sort(Sv_u,Sv),
	each_body_succ_builtin_(Call,AbsInt,T,Tg,Vs,SgKey,Sg,Sv,HvFv_u,Succ).
body_succ0(_RFlag,SgKey,Sg,Sv_u,HvFv_u,Call,Succ,_List0,_List,AbsInt,
	   _ClId,_F,_N,_Id):-
	sort(Sv_u,Sv),
	each_call_to_success(Call,SgKey,Sg,Sv,HvFv_u,AbsInt,Succ).

%% predicate adapted from fixpo_ops
each_body_succ_builtin_([],_,_T,_Tg,_,_,_Sg,_Sv,_HvFv_u,[]).
each_body_succ_builtin_([Call|Calls],AbsInt,T,Tg,Vs,SgKey,Sg,Sv,HvFv_u,[Succ|Succs]):-
	project(AbsInt,Sg,Sv,HvFv_u,Call,Proj),
	body_succ_builtin(T,AbsInt,Tg,Vs,Sv,HvFv_u,Call,Proj,Succ),!, %% Doamin call
	each_body_succ_builtin_tabling_(Calls,AbsInt,T,Tg,Vs,SgKey,Sg,Sv,HvFv_u,Succs).

each_call_to_success([Call],SgKey,Sg,Sv,HvFv_u,AbsInt,Succ):-
	!,
	project(AbsInt,Sg,Sv,HvFv_u,Call,Proj),
	call_to_success(SgKey,Call,Proj,Sg,Sv,AbsInt,Succ).

each_call_to_success(LCall,SgKey,Sg,Sv,HvFv_u,AbsInt,LSucc):-
	each_call_to_success0(LCall,SgKey,Sg,Sv,HvFv_u,AbsInt,
                              LSucc).

each_call_to_success0([],_SgK,_Sg,_Sv,_HvFv,_AbsInt,[]).
each_call_to_success0([Call|LCall],SgKey,Sg,Sv,HvFv_u,AbsInt,
	              LSucc):-
	project(AbsInt,Sg,Sv,HvFv_u,Call,Proj),
	call_to_success(SgKey,Call,Proj,Sg,Sv,AbsInt,LSucc0),
	append(LSucc0,LSucc1,LSucc),
	each_call_to_success0(LCall,SgKey,Sg,Sv,HvFv_u,AbsInt,
	                      LSucc1).

meta_call([],_HvFv_u,Call,[],Call,List,List,_AbsInt,_ClId,_Id,[]).
meta_call([Body|Bodies],HvFv_u,Call,Succ0,Succ,L0,List,AbsInt,ClId,Id,Ids):-
	meta_call_([Body|Bodies],HvFv_u,Call,Succ0,Succ,L0,List,AbsInt,ClId,Id,Ids).
meta_call_([Body|Bodies],HvFv_u,Call,Succ0,Succ,L0,List,AbsInt,ClId,Id,Ids):-
	meta_call_body(Body,ClId,Call,Succ1,L0,L1,HvFv_u,AbsInt,Id,Ids0),
	widen_succ(AbsInt,Succ0,Succ1,Succ2),
	append(Succ0,Succ1,Succ2),
	append(Ids0,Ids1,Ids),
	meta_call_(Bodies,HvFv_u,Call,Succ2,Succ,L1,List,AbsInt,ClId,Id,Ids1).
meta_call_([],_HvFv_u,_Call,Succ,Succ,List,List,_AbsInt,_ClId,_Id,[]).

meta_call_body((Sg,Rest),K,Call,Exit,OldList,NewList,Vars_u,AbsInt,PId,CIds):-
	!,
	CIds=[Id|Ids],
	body_succ(Call,Sg,Succ,OldList,IntList,Vars_u,AbsInt,K,PId,Id),
	meta_call_body(Rest,K,Succ,Exit,IntList,NewList,Vars_u,AbsInt,PId,Ids).
meta_call_body(true,_,Call,Call,List,List,_,_,_,[no]):- !.
meta_call_body(Sg,Key,Call,Exit,OldList,NewList,Vars_u,AbsInt,PId,[Id]):- 
	body_succ(Call,Sg,Exit,OldList,NewList,Vars_u,AbsInt,Key,PId,Id).

concretes_to_body([],_SgKey,_AbsInt,[]).
concretes_to_body([Sg|Sgs],SgKey,AbsInt,[B|Bs]):-
	body_info0(Sg:SgKey,[],AbsInt,B),
	concretes_to_body(Sgs,SgKey,AbsInt,Bs).

%------------------------------------------------------------------------%
% query(+,+,+,+,+,+,+,-)                                                 %
%------------------------------------------------------------------------%

:- doc(query(AbsInt,QKey,Query,Qv,RFlag,N,Call,Succ),
	"The success pattern of @var{Query} with @var{Call} is
	@var{Succ} in the analysis domain @var{AbsInt}. The predicate
	called is identified by @var{QKey}. The goal @var{Query} has
        variables @var{Qv}.").

query(AbsInt,QKey,Query,Qv,_RFlag,_N,Call,Succ) :-
	project(AbsInt,Query,Qv,Qv,Call,Proj),
	call_to_success(QKey,Call,Proj,Query,Qv,AbsInt,Succ), !.

query(_AbsInt,_QKey,_Query,_Qv,_RFlag,_N,_Call,_Succ):-
	% should never happen, but...
	error_message("SOMETHING HAS FAILED!").
\end{lstlisting}

\vspace{1em}

\section{PLAI Algorithm Implementation in Ciao Prolog}
\label{sec:Ap-plai-CiaoPP}

We include here the Ciao Prolog implementation of PLAI.  As mentioned
before, we have removed the comments from the file since the goal of
this appendix it to make it easier for the reader to compare the Ciao
Prolog code w.r.t.\ the code using TCLP, which we include
in~\ref{sec:Ap-plai-TCLP}.  The original version is available at
\papermaterial.

\vspace{1em}

\begin{lstlisting}[style=MyASP, mathescape = false, basicstyle =
\footnotesize\ttfamily\color{PrologPredicate}]
/*             Copyright (C)1990-2019 UPM-CLIP				*/

:- module(fixpo_plai_with_comments,
	[ query/8,
	  init_fixpoint/0,
	  cleanup_fixpoint/1,
	  entry_to_exit/9
	],
	[assertions, datafacts]).

% Ciao library
:- use_module(library(aggregates), [bagof/3, (^)/2]).
:- use_module(library(lists), [member/2, append/3]).
:- use_module(library(terms_vars), [varset/2]).
:- use_module(library(sets), [merge/3, ord_subtract/3]).
:- use_module(library(sort), [sort/2]).
:- use_module(library(messages)).

% CiaoPP library
:- use_module(ciaopp(preprocess_flags), [current_pp_flag/2, set_pp_flag/2]).

% Plai library
:- use_module(ciaopp(plai/fixpo_ops), [inexistent/2, variable/2, bottom/1, 
	singleton/2, fixpoint_id_reuse_prev/5, fixpoint_id/1, fixp_id/1,
	each_abs_sort/3,
	each_extend/6, each_project/6, each_exit_to_prime/8, each_unknown_call/4,
	each_body_succ_builtin/12, body_succ_meta/7, reduce_equivalent/3,
	each_apply_trusted/7,	widen_succ/4,	decide_memo/6,clause_applies/2,
	abs_subset_/3]).

:- use_module(ciaopp(plai/domains)).
:- use_module(ciaopp(plai/trace_fixp), [fixpoint_trace/7, cleanup/0]).
:- use_module(ciaopp(plai/plai_db), 
	[ complete/7, memo_call/5, memo_table/6, cleanup_plai_db/1, patch_parents/6 ]).
:- use_module(ciaopp(plai/psets), [update_if_member_idlist/3]).
:- use_module(ciaopp(plai/re_analysis), [erase_previous_memo_tables_and_parents/4]).
:- use_module(ciaopp(plai/transform), [body_info0/4, trans_clause/3]).
:- use_module(ciaopp(plai/apply_assertions_old),
	[ apply_trusted0/7, 
 	  cleanup_trusts/1 ]).

:- doc(author,"Kalyan Muthukumar").
:- doc(author,"Maria Garcia de la Banda").
:- doc(author,"Francisco Bueno").

:- doc(module,"This module implements the top-down fixpoint 
	algorithm of PLAI, both in its mono-variant and multi-variant
	on successes versions. It is always multi-variant on calls.
        The algorithm is parametric on the particular analysis domain.").


:- data '$depend_list'/3.
:- data ch_id/2.

:- data approx/6.
:- data fixpoint/6.
:- data fixpoint_variant/6.
:- data approx_variant/7.

init_fixpoint:-
	retractall_fact(approx(_,_,_,_,_,_)),
	retractall_fact(fixpoint(_,_,_,_,_,_)),
	retractall_fact('$depend_list'(_,_,_)),
	retractall_fact(ch_id(_,_)),
	retractall_fact(fixpoint_variant(_,_,_,_,_,_)),
	retractall_fact(approx_variant(_,_,_,_,_,_,_)),
	trace_fixp:cleanup.

cleanup_fixpoint(AbsInt):-
	cleanup_plai_db(AbsInt),
	cleanup_trusts(AbsInt),
	retractall_fact(fixp_id(_)),
	asserta_fact(fixp_id(0)), % there is no way to recover this 
	init_fixpoint.            % if several analyses coexist!

approx_to_completes(AbsInt):-
	current_fact(approx(SgKey,Sg,Proj,Prime,Pid,Fs),Ref),
	asserta_fact(complete(SgKey,AbsInt,Sg,Proj,Prime,Pid,Fs)),
	erase(Ref),
	fail.
approx_to_completes(AbsInt):-
	current_fact(approx_variant(_Id,Pid,SgKey,Sg,Proj,Prime,Fs),Ref),	
	asserta_fact(complete(SgKey,AbsInt,Sg,Proj,Prime,Pid,Fs)),
	erase(Ref),
	fail.
approx_to_completes(_AbsInt).


%------------------------------------------------------------------------%
% call_to_success(+,+,+,+,+,+,+,-,-,+,+,+)                               %
%------------------------------------------------------------------------%

call_to_success(_RFlag,SgKey,Call,Proj,Sg,Sv,AbsInt,_ClId,Succ,List,F,N,Id):-
	% ClId = number identifying the clause?... for an entry point is 0...
	% F = program point of the call. clauseId+/0 for an entry call
	current_fact(complete(SgKey,AbsInt,Subg,Proj1,Prime1,Id,Fs),R),
	identical_proj(AbsInt,Sg,Proj,Subg,Proj1), !,
	patch_parents(R,complete(SgKey,AbsInt,Subg,Proj1,Prime1,Id,Ps),F,N,Ps,Fs),
	List = [],
	each_abs_sort(Prime1,AbsInt,Prime),
	each_extend(Sg,Prime,AbsInt,Sv,Call,Succ).
call_to_success(r,SgKey,Call,Proj,Sg,Sv,AbsInt,_ClId,Succ,List,F,N,Id) :-
	current_fact(approx(SgKey,Subg,Proj1,Prime1,Id,Fs),Ref),
	identical_proj(AbsInt,Sg,Proj,Subg,Proj1), !,
	each_abs_sort(Prime1,AbsInt,TempPrime),
	current_fact('$depend_list'(Id,SgKey,IdList)),
	call_to_success_approx(SgKey,Subg,Call,Proj,Proj1,Sg,Sv,AbsInt,F,N,Fs,
	                       Id,Ref,IdList,Prime1,TempPrime,List,Prime),
	each_extend(Sg,Prime,AbsInt,Sv,Call,Succ).
call_to_success(r,SgKey,Call,Proj,Sg,Sv,AbsInt,_ClId,Succ,List,F,N,Id):-
	current_fact(fixpoint(SgKey,Subg,Proj1,Prime1,Id,Fs),Ref),
	identical_proj(AbsInt,Sg,Proj,Subg,Proj1), !,
	patch_parents(Ref,fixpoint(SgKey,Subg,Proj1,Prime1,Id,Ps),F,N,Ps,Fs),
	current_fact(ch_id(Id,Num)),
	List = [Id/Num],
	each_abs_sort(Prime1,AbsInt,Prime),
	each_extend(Sg,Prime,AbsInt,Sv,Call,Succ).
call_to_success(_RFlag,SgKey,Call,Proj,Sg,Sv,AbsInt,_ClId,Succ,List,F,N,Id):-
	current_pp_flag(variants,on),
	current_fact(complete(SgKey,AbsInt,Subg,Proj1,Prime1,_Id1,_Fs),_R),
	identical_proj_1(AbsInt,Sg,Proj,Subg,Proj1,Prime1,Prime2), !,
	format("call to success tipe _RFlag SgKey",[]),
	&( current_pp_flag(reuse_fixp_id,on) ->
	    fixpoint_id_reuse_prev(SgKey,AbsInt,Sg,Proj,Id)
	;
	    fixpoint_id(Id)
	&),
	each_abs_sort(Prime2,AbsInt,Prime),
	List = [],
	asserta_fact(complete(SgKey,AbsInt,Sg,Proj,Prime,Id,[(F,N)])),
	each_extend(Sg,Prime,AbsInt,Sv,Call,Succ).
call_to_success(r,SgKey,Call,Proj,Sg,Sv,AbsInt,_ClId,Succ,List,F,N,Id) :-
	current_pp_flag(variants,on),
	current_fact(approx(SgKey,Subg,Proj1,Prime1,Id1,Fs),Ref),
	identical_proj_1(AbsInt,Sg,Proj,Subg,Proj1,Prime1,Prime2), !,
	each_abs_sort(Prime2,AbsInt,TempPrime),
	current_fact('$depend_list'(Id1,SgKey,IdList)),
	call_to_success_approx_variant(SgKey,Subg,Call,Proj,Proj1,Sg,Sv,AbsInt,F,N,Fs,
	                       Id,Id1,Ref,IdList,Prime1,TempPrime,List,Prime),
	each_extend(Sg,Prime,AbsInt,Sv,Call,Succ).
call_to_success(r,SgKey,Call,Proj,Sg,Sv,AbsInt,_ClId,Succ,List,F,N,Id):-
	current_pp_flag(variants,on),
	current_fact(fixpoint(SgKey,Subg,Proj1,Prime1,Id1,_Fs),_Ref),
	identical_proj_1(AbsInt,Sg,Proj,Subg,Proj1,Prime1,Prime2), !,
	&(
	    current_fact(fixpoint_variant(Id1,Id,SgKey,Sgv,Projv,Fsv),Refv),
	    identical_proj(AbsInt,Sg,Proj,Sgv,Projv) ->
	    patch_parents(Refv,fixpoint_variant(Id1,Id,SgKey,Sgv,Projv,Ps),F,N,Ps,Fsv)
	;
	    &(
		current_pp_flag(reuse_fixp_id,on) ->
		fixpoint_id_reuse_prev(SgKey,AbsInt,Sg,Proj,Id)
	    ;
		fixpoint_id(Id)
	    &),
	    asserta_fact(fixpoint_variant(Id1,Id,SgKey,Sg,Proj,[(F,N)]))
	&),
	each_abs_sort(Prime2,AbsInt,Prime),
	current_fact(ch_id(Id1,Num)),
	List = [Id1/Num],
	each_extend(Sg,Prime,AbsInt,Sv,Call,Succ).
call_to_success(r,SgKey,Call,Proj,Sg,Sv,AbsInt,_ClId,Succ,List,F,N,Id) :-
	init_fixpoint0(SgKey,Call,Proj,Sg,Sv,AbsInt,F,N,[(F,N)],Id,List,Prime),
	each_extend(Sg,Prime,AbsInt,Sv,Call,Succ).
call_to_success(nr,SgKey,Call,Proj,Sg,Sv,AbsInt,ClId,Succ,[],F,N,Id):-
	&( current_pp_flag(reuse_fixp_id,on) ->
	    fixpoint_id_reuse_prev(SgKey,AbsInt,Sg,Proj,Id)
	;
	    fixpoint_id(Id)
	&),
	proj_to_prime_nr(SgKey,Sg,Sv,Call,Proj,AbsInt,ClId,Prime,Id),
	asserta_fact(complete(SgKey,AbsInt,Sg,Proj,Prime,Id,[(F,N)])),
	each_extend(Sg,Prime,AbsInt,Sv,Call,Succ).

call_to_success_approx(SgKey,Subg,_Call,Proj,Proj1,Sg,_Sv,_AbsInt,F,N,Fs,
	                     Id,Ref,IdList,Prime1,TempPrime,List,Prime):-
	not_modified(IdList), !,
	patch_parents(Ref,approx(SgKey,Subg,Proj1,Prime1,Id,Ps),F,N,Ps,Fs),
	Prime = TempPrime,
	List = IdList.
call_to_success_approx(SgKey,_Subg,Call,Proj,_Proj1,Sg,Sv,AbsInt,F,N,Fs,
	                     Id,Ref,_IdList,_Prime1,TempPrime,List,Prime):-
	erase(Ref),
	init_fixpoint_(SgKey,Call,Proj,Sg,Sv,AbsInt,F,N,Fs,Id,
	                     TempPrime,List,Prime).

aproxs_to_fixpoint_variant(Id):-
	current_fact(approx_variant(Id,Idv,SgKey,Sgv,Projv,_Primev,Fs),Ref),!,
	erase(Ref),
	asserta_fact(fixpoint_variant(Id,Idv,SgKey,Sgv,Projv,Fs)),
	aproxs_to_fixpoint_variant(Id).
aproxs_to_fixpoint_variant(_).


call_to_success_approx_variant(SgKey,_Subg,_Call,Proj,_Proj1,Sg,_Sv,AbsInt,F,N,_Fs,
	                     Id,Id1,_Ref,IdList,_Prime1,TempPrime,List,Prime):-
	not_modified(IdList), !,
	&(
	    current_fact(approx_variant(Id1,Id,SgKey,Sgv,Projv,Primev,Fsv),Refv),
	    identical_proj(AbsInt,Sg,Proj,Sgv,Projv) ->
	    patch_parents(Refv,approx_variant(Id1,Id,SgKey,Sgv,Projv,Primev,Ps),F,N,Ps,Fsv) 
	;
	    &(
		current_pp_flag(reuse_fixp_id,on) ->
		fixpoint_id_reuse_prev(SgKey,AbsInt,Sg,Proj,Id)
	    ;
		fixpoint_id(Id)
	    &),
	    asserta_fact(approx_variant(Id1,Id,SgKey,Sg,Proj,TempPrime,[(F,N)]))
	&),
	Prime = TempPrime,
	List = IdList.
call_to_success_approx_variant(SgKey,Subg,Call,Proj,Proj1,Sg,Sv,AbsInt,F,N,Fs,
	                     Id,Id1,Ref,_IdList,Prime1,_TempPrime,List,Prime):-
	&(
	    current_fact(approx_variant(Id1,Id,SgKey,Sgv,Projv,_Primev,Fsv),Refv),
	    identical_proj(AbsInt,Sg,Proj,Sgv,Projv) ->
	    erase(Refv),
	    &( member((F,N),Fsv) -> NewFs = Fsv ; NewFs = [(F,N)|Fsv] %)
	;
	    &(
		current_pp_flag(reuse_fixp_id,on) ->
		fixpoint_id_reuse_prev(SgKey,AbsInt,Sg,Proj,Id)
	    ;
		fixpoint_id(Id)
	    &),
	    NewFs = [(F,N)]
	&),
        aproxs_to_fixpoint_variant(Id1),
	erase(Ref),
	asserta_fact(fixpoint_variant(Id1,Id,SgKey,Sg,Proj,NewFs)),
	varset(Subg,Subv),
	init_fixpoint_(SgKey,Call,Proj1,Subg,Subv,AbsInt,F,N,Fs,Id1,
	                     Prime1,List,Prime0),
	each_exit_to_prime(Prime0,AbsInt,Sg,Subv,Subg,Sv,(no,Proj),Prime).

init_fixpoint0(SgKey,Call,Proj,Sg,Sv,AbsInt,F,N,Fs,Id,List,Prime):-
	init_fixpoint2(SgKey,Call,Proj,Sg,Sv,AbsInt,F,N,Fs,Id,List,Prime).

init_fixpoint1(SgKey,_Call,Proj,Sg,_Sv,AbsInt,F,N,_Fs0,Id,List,Prime):-
	current_fact(complete(SgKey,AbsInt,Subg,Proj1,Prime1,Id,Fs),R),
	identical_proj(AbsInt,Sg,Proj,Subg,Proj1), !,
	patch_parents(R,complete(SgKey,AbsInt,Subg,Proj1,Prime1,Id,Ps),F,N,Ps,Fs),
	List = [],
	each_abs_sort(Prime1,AbsInt,Prime).
init_fixpoint1(SgKey,Call,Proj,Sg,Sv,AbsInt,F,N,_Fs0,Id,List,Prime):-
	current_fact(approx(SgKey,Subg,Proj1,Prime1,Id,Fs),Ref),
	identical_proj(AbsInt,Sg,Proj,Subg,Proj1), !,
	each_abs_sort(Prime1,AbsInt,TempPrime),
	current_fact('$depend_list'(Id,SgKey,IdList)),
	call_to_success_approx(SgKey,Subg,Call,Proj,Proj1,Sg,Sv,AbsInt,F,N,Fs,
	                       Id,Ref,IdList,Prime1,TempPrime,List,Prime).
init_fixpoint1(SgKey,_,Proj,Sg,_Sv,AbsInt,F,N,_Fs0,Id,List,Prime):-
	current_fact(fixpoint(SgKey,Subg,Proj1,Prime1,Id,Fs),Ref),
	identical_proj(AbsInt,Sg,Proj,Subg,Proj1), !,
	patch_parents(Ref,fixpoint(SgKey,Subg,Proj1,Prime1,Id,Ps),F,N,Ps,Fs),
	current_fact(ch_id(Id,Num)),
	List = [Id/Num],
	each_abs_sort(Prime1,AbsInt,Prime).
init_fixpoint1(SgKey,_Call,Proj,Sg,_Sv,AbsInt,F,N,_Fs0,Id,List,Prime):-
	current_pp_flag(variants,on),
	current_fact(complete(SgKey,AbsInt,Subg,Proj1,Prime1,_Id1,_Fs),_R),
	identical_proj_1(AbsInt,Sg,Proj,Subg,Proj1,Prime1,Prime2), !,
	&( current_pp_flag(reuse_fixp_id,on) ->
	    fixpoint_id_reuse_prev(SgKey,AbsInt,Sg,Proj,Id)
	;
	    fixpoint_id(Id)
	&),
	each_abs_sort(Prime2,AbsInt,Prime),
	List = [],
	asserta_fact(complete(SgKey,AbsInt,Sg,Proj,Prime,Id,[(F,N)])).
init_fixpoint1(SgKey,Call,Proj,Sg,Sv,AbsInt,F,N,_Fs0,Id,List,Prime):-
	current_pp_flag(variants,on),
	current_fact(approx(SgKey,Subg,Proj1,Prime1,Id1,Fs),Ref),
	identical_proj_1(AbsInt,Sg,Proj,Subg,Proj1,Prime1,Prime2), !,
	each_abs_sort(Prime2,AbsInt,TempPrime),
	current_fact('$depend_list'(Id1,SgKey,IdList)),
	call_to_success_approx_variant(SgKey,Subg,Call,Proj,Proj1,Sg,Sv,AbsInt,F,N,Fs,
	                       Id,Id1,Ref,IdList,Prime1,TempPrime,List,Prime).
init_fixpoint1(SgKey,_,Proj,Sg,_Sv,AbsInt,F,N,_Fs0,Id,List,Prime):-
	current_pp_flag(variants,on),
	current_fact(fixpoint(SgKey,Subg,Proj1,Prime1,Id1,_Fs),_Ref),
	identical_proj_1(AbsInt,Sg,Proj,Subg,Proj1,Prime1,Prime2), !,
	&(
	    current_fact(fixpoint_variant(Id1,Id,SgKey,Sgv,Projv,Fsv),Refv),
	    identical_proj(AbsInt,Sg,Proj,Sgv,Projv) ->
	    patch_parents(Refv,fixpoint_variant(Id1,Id,SgKey,Sgv,Projv,Ps),F,N,Ps,Fsv)
	;
	    &(
		current_pp_flag(reuse_fixp_id,on) ->
		fixpoint_id_reuse_prev(SgKey,AbsInt,Sg,Proj,Id)
	    ;
		fixpoint_id(Id)
	    &),
	    asserta_fact(fixpoint_variant(Id1,Id,SgKey,Sg,Proj,[(F,N)]))
	&),
	each_abs_sort(Prime2,AbsInt,Prime),
	current_fact(ch_id(Id1,Num)),
	List = [Id1/Num].
init_fixpoint1(SgKey,Call,Proj,Sg,Sv,AbsInt,F,N,Fs,Id,List,Prime):-
	init_fixpoint2(SgKey,Call,Proj,Sg,Sv,AbsInt,F,N,Fs,Id,List,Prime).

init_fixpoint2(SgKey,Call,Proj,Sg,Sv,AbsInt,F,N,Fs,Id,List,Prime):-
	&( current_pp_flag(reuse_fixp_id,on) ->
	    fixpoint_id_reuse_prev(SgKey,AbsInt,Sg,Proj,Id)
	;
	    fixpoint_id(Id)
	&),
	asserta_fact(ch_id(Id,1)),
	proj_to_prime_r(SgKey,Sg,Sv,Call,Proj,AbsInt,TempPrime,Id), 
	init_fixpoint_(SgKey,Call,Proj,Sg,Sv,AbsInt,F,N,Fs,Id,
	                     TempPrime,List,Prime).

init_fixpoint_(SgKey,Call,Proj,Sg,Sv,AbsInt,F,N,Fs,Id,Prime0,List,Prime):-
	normalize_asub0(AbsInt,Prime0,TempPrime),
	asserta_fact(fixpoint(SgKey,Sg,Proj,TempPrime,Id,Fs)),
	bagof(X, X^(trans_clause(SgKey,r,X)),Clauses),!,
	fixpoint_compute(Clauses,SgKey,Sg,Sv,Call,Proj,
	                    AbsInt,_LEntry,TempPrime,Prime1,Id,TempList),
	each_apply_trusted(Proj,SgKey,Sg,Sv,AbsInt,Prime1,Prime),
	current_fact(fixpoint(SgKey,Sg,_,_,Id,Fs2),Ref),
	erase(Ref),
	&( current_fact('$depend_list'(Id,SgKey,_),RefDep) ->
	  erase(RefDep)
	; true
	&),
	update_if_member_idlist(TempList,Id,AddList),
	( member((F,N),Fs2) -> NewFs = Fs2 ; NewFs = [(F,N)|Fs2] ),
	decide_approx(AddList,Id,NewFs,AbsInt,SgKey,Sg,Proj,Prime),
	List = AddList.

widen_call(AbsInt,SgKey,Sg,F1,Id0,Proj1,Proj):-
	&( current_pp_flag(widencall,off) -> fail ; true &),
	widen_call0(AbsInt,SgKey,Sg,F1,Id0,[Id0],Proj1,Proj), !.

widen_call0(AbsInt,SgKey,Sg,F1,Id0,Ids,Proj1,Proj):-
	widen_call1(AbsInt,SgKey,Sg,F1,Id0,Ids,Proj1,Proj).
widen_call0(AbsInt,SgKey,Sg,F1,Id0,Ids,Proj1,Proj):-
	current_pp_flag(widencall,com_child),
	widen_call2(AbsInt,SgKey,Sg,F1,Id0,Ids,Proj1,Proj).

widen_call1(AbsInt,SgKey,Sg,F1,Id0,Ids,Proj1,Proj):-
	current_fact(fixpoint(SgKey0,Sg0,Proj0,_Prime0,Id0,Fs0)),
	&( SgKey=SgKey0,
          % same program point:
	  member((F1,_NewId0),Fs0)
	-> Sg0=Sg,
	   abs_sort(AbsInt,Proj0,Proj0_s),
	   abs_sort(AbsInt,Proj1,Proj1_s),
	   widencall(AbsInt,Proj0_s,Proj1_s,Proj)
	 ; % continue with the parents:
	   member((_F1,NewId0),Fs0),
	   \+ member(NewId0,Ids),
	   widen_call1(AbsInt,SgKey,Sg,F1,NewId0,[NewId0|Ids],Proj1,Proj)
	&).

widen_call2(AbsInt,SgKey,Sg,F1,_Id,_Ids,Proj1,Proj):-
	current_fact(complete(SgKey,AbsInt,Sg0,Proj0,_Prime0,_Id0,Fs0)),
	member((F1,Id0),Fs0),
	Sg0=Sg,
	same_fixpoint_ancestor(Id0,[Id0],AbsInt),
	abs_sort(AbsInt,Proj0,Proj0_s),
	abs_sort(AbsInt,Proj1,Proj1_s),
	widencall(AbsInt,Proj0_s,Proj1_s,Proj).

same_fixpoint_ancestor(Id0,_Ids,_AbsInt):-
	current_fact(fixpoint(_SgKey0,_Sg0,_Proj0,_Prime0,Id0,_Fs0)), !.
same_fixpoint_ancestor(Id0,_Ids,_AbsInt):-
	current_fact(approx(_SgKey0,_Sg0,_Proj0,_Prime0,Id0,_Fs0)), !.
same_fixpoint_ancestor(Id0,Ids,AbsInt):-
	current_fact(complete(_SgKey0,AbsInt,_Sg0,_Proj0,_Prime0,Id0,Fs0)),
	member((_F1,Id),Fs0),
	\+ member(Id,Ids),
	same_fixpoint_ancestor(Id,[Id|Ids],AbsInt).

fixpoint_variants_update(Id,AbsInt,Sg,Prime):-
	current_fact(fixpoint_variant(Id,Idv,SgKey,Sgv,Projv,Fs),Ref),!,
	erase(Ref),
	varset(Sg,Hv),
	varset(Sgv,Hvv),
	each_exit_to_prime(Prime,AbsInt,Sgv,Hv,Sg,Hvv,(no,Projv),Prime2),
	asserta_fact(complete(SgKey,AbsInt,Sgv,Projv,Prime2,Idv,Fs)),	
	fixpoint_variants_update(Id,AbsInt,Sg,Prime). 
fixpoint_variants_update(_,_,_,_).

approx_variants_update(Id,AbsInt,Sg,Prime):-
	current_fact(fixpoint_variant(Id,Idv,SgKey,Sgv,Projv,Fs),Ref),!,
	erase(Ref),
	varset(Sg,Hv),
	varset(Sgv,Hvv),
	each_exit_to_prime(Prime,AbsInt,Sgv,Hv,Sg,Hvv,(no,Projv),Prime2),
	asserta_fact(approx_variant(Id,Idv,SgKey,Sgv,Projv,Prime2,Fs)),	
	approx_variants_update(Id,AbsInt,Sg,Prime). 
approx_variants_update(_,_,_,_).
	
decide_approx([],Id,Fs,AbsInt,SgKey,Sg,Proj,Prime):- !,
	current_fact(ch_id(Id,_),Ref3),
	erase(Ref3),
        % Not needed for correctness: only book-keeping
	% update_depend_list_approx(Id,AbsInt),
	asserta_fact(complete(SgKey,AbsInt,Sg,Proj,Prime,Id,Fs)),
	&(
	    current_pp_flag(variants,on) -> 
	    each_abs_sort(Prime,AbsInt,Prime_s),
	    fixpoint_variants_update(Id,AbsInt,Sg,Prime_s)
	;
	    true
	&).
decide_approx(AddList,Id,Fs,_AbsInt,SgKey,Sg,Proj,Prime):-
	asserta_fact('$depend_list'(Id,SgKey,AddList)),
	asserta_fact(approx(SgKey,Sg,Proj,Prime,Id,Fs),_),
	&(
	    current_pp_flag(variants,on) -> 
	    each_abs_sort(Prime,AbsInt,Prime_s),
	    approx_variants_update(Id,AbsInt,Sg,Prime_s)
	;
	    true
	&).

not_modified([]).
not_modified([Id/N|List]):-
	current_fact(ch_id(Id,N)), !,
	not_modified(List).

proj_to_prime_nr(SgKey,Sg,Sv,Call,Proj,AbsInt,_ClId,LPrime,Id) :-
	bagof(X, X^(trans_clause(SgKey,nr,X)),Clauses), !,
	proj_to_prime(Clauses,SgKey,Sg,Sv,Call,Proj,AbsInt,LPrime1,Id),
	compute_clauses_lub(AbsInt,Proj,LPrime1,LPrime).
proj_to_prime_nr(SgKey,Sg,Sv,_Call,Proj,AbsInt,ClId,LPrime,_Id) :-
	apply_trusted0(Proj,SgKey,Sg,Sv,AbsInt,ClId,Prime), !,
	singleton(Prime,LPrime).
proj_to_prime_nr(_SgKey,Sg,Sv,Call,_Proj,AbsInt,_ClId,LSucc,_Id) :-
	% In Java programs, mode and type information is known for any method.
	% Therefore, in case of a method with unavailable code we can still
	% infer useful information.
	&( current_pp_flag(prog_lang,java) ->
	  unknown_call(AbsInt,Sg,Sv,Call,Succ),
	  singleton(Succ,LSucc)
	;
	  fail
	&).
proj_to_prime_nr(SgKey,_Sg,_Sv,_Call,_Proj,_AbsInt,ClId,Bot,_Id) :-
	bottom(Bot),
	inexistent(SgKey,ClId).

proj_to_prime_r(SgKey,Sg,Sv,Call,Proj,AbsInt,Prime,Id) :-
	bagof(X, X^(trans_clause(SgKey,nr,X)),Clauses), !,
	proj_to_prime(Clauses,SgKey,Sg,Sv,Call,Proj,AbsInt,Prime,Id).
proj_to_prime_r(_SgKey,_Sg,_Sv,_Call,_Proj,_AbsInt,Bot,_Id):-
	bottom(Bot).

proj_to_prime(Clauses,SgKey,Sg,Sv,Call,Proj,AbsInt,Prime,Id) :-
	proj_to_prime_loop(Clauses,Sg,Sv,Call,Proj,AbsInt,ListPrime0,Id),
	reduce_equivalent(ListPrime0,AbsInt,ListPrime1),
	each_apply_trusted(Proj,SgKey,Sg,Sv,AbsInt,ListPrime1,Prime).

proj_to_prime_loop([],_,_,_,_,_,[],_).
proj_to_prime_loop([Clause|Rest],Sg,Sv,Call,Proj,AbsInt,Primes,Id):-
	do_nr_cl(Clause,Sg,Sv,Call,Proj,AbsInt,Primes,TailPrimes,Id),!,
	proj_to_prime_loop(Rest,Sg,Sv,Call,Proj,AbsInt,TailPrimes,Id).

do_nr_cl(Clause,Sg,Sv,Call,Proj,AbsInt,Primes,TailPrimes,Id):-
	Clause = clause(Head,Vars_u,K,Body),
	clause_applies(Head,Sg), !,
	varset(Head,Hv),
	sort(Vars_u,Vars),
	ord_subtract(Vars,Hv,Fv),
	process_body(Body,K,AbsInt,Sg,Hv,Fv,Vars_u,Head,Sv,Call,
	                                                Proj,LPrime,Id),
	append_(LPrime,TailPrimes,Primes).
do_nr_cl(_Clause,_Sg,_Sv,_Call,_Proj,_AbsInt,Primes,Primes,_Id).

append_([Prime],TailPrimes,Primes):- !, Primes=[Prime|TailPrimes].
append_(LPrime,TailPrimes,Primes):- append(LPrime,TailPrimes,Primes).

process_body(Body,K,AbsInt,Sg,Hv,Fv,_,Head,Sv,Call,Proj,LPrime,Id):- 
	Body = g(_,[],'$built'(_,true,_),'true/0',true), !,
	Help=(Sv,Sg,Hv,Fv,AbsInt),
	singleton(Prime,LPrime),
	call_to_success_fact(AbsInt,Sg,Hv,Head,K,Sv,Call,Proj,Prime,_Succ),
	&( current_pp_flag(fact_info,on) -> 
 	  call_to_entry(AbsInt,Sv,Sg,Hv,Head,K,[],Prime,Exit,_),
	  decide_memo(AbsInt,K,Id,no,Hv,[Exit])
	;
	  true
	&).
process_body(Body,K,AbsInt,Sg,Hv,Fv,Vars_u,Head,Sv,_,Proj,Prime,Id):-
	call_to_entry(AbsInt,Sv,Sg,Hv,Head,K,Fv,Proj,Entry,ExtraInfo),
	singleton(Entry,LEntry),
	entry_to_exit(Body,K,LEntry,Exit,[],_,Vars_u,AbsInt,Id),
	each_exit_to_prime(Exit,AbsInt,Sg,Hv,Head,Sv,ExtraInfo,Prime).

fixpoint_compute(Clauses,SgKey,Sg,Sv,Call,Proj,AbsInt,LEntryInf,
	         Prime0,Prime,Id,List) :-
	fixpoint_compute_(Clauses,SgKey,Sg,Sv,Call,Proj,AbsInt,LEntryInf,
	                  Prime0,Prime1,Id,List),
	compute_clauses_lub(AbsInt,Proj,Prime1,Prime).

fixpoint_compute_(Clauses,SgKey,Sg,Sv,Call,Proj,AbsInt,LEntryInf,
	         TempPrime,Prime,Id,List) :-
        compute(Clauses,SgKey,Sg,Sv,Call,Proj,AbsInt,LEntryInf,
	         TempPrime,Prime1,Id,[],NewList,Flag),
	fixpoint(NewList,Flag,Clauses,SgKey,Sg,Sv,Call,Proj,AbsInt,LEntryInf,
	         Prime1,Prime,Id,List), !.

fixpoint([],_,_,_,_,_,_,_,_,_,Prime1,Prime,_,List):- !,
	Prime = Prime1,
	List = [].
fixpoint(NewList,Flag,_,_,_,_,_,_,_,_,Prime1,Prime,_,List):- 
        var(Flag),!,
	Prime = Prime1,
	List = NewList.
fixpoint(_,_,Clauses,SgKey,Sg,Sv,Call,Proj,AbsInt,LEntryInf,Prime1,Prime,Id,List):-
        fixpoint_compute_(Clauses,SgKey,Sg,Sv,Call,Proj,AbsInt,LEntryInf,
	         Prime1,Prime,Id,List).

% some domains need normalization to perform the widening:
normalize_asub0(AbsInt,Prime0,Prime):-
	current_pp_flag(widen,on), !,
	normalize_asub(AbsInt,Prime0,Prime).
normalize_asub0(_AbsInt,Prime,Prime).

compute([],_,_,_,_,_,_,[],Prime,Prime,_,List,List,_).
compute([Clause|Rest],SgKey,Sg,Sv,Call,Proj,AbsInt,[EntryInf|LEntryInf],
	                  TempPrime,Prime,Id,List,NewList,Flag) :-
	do_r_cl(Clause,SgKey,Sg,Sv,Proj,AbsInt,EntryInf,Id,List,IntList,
	                                   TempPrime,NewPrime,Flag),
	compute(Rest,SgKey,Sg,Sv,Call,Proj,AbsInt,LEntryInf,NewPrime,Prime,
	                                   Id,IntList,NewList,Flag).

do_r_cl(Clause,SgKey,Sg,Sv,Proj,AbsInt,EntryInf,Id,OldL,List,TempPrime,
	                                               NewPrime,Flag):-
	Clause=clause(Head,Vars_u,K,Body),
	clause_applies(Head,Sg), !,
	erase_previous_memo_tables_and_parents(Body,AbsInt,K,Id),
	varset(Head,Hv),
	reuse_entry(EntryInf,Vars_u,AbsInt,Sv,Sg,Hv,Head,K,Proj,Entry,ExtraInfo),
	singleton(Entry,LEntry),
	entry_to_exit(Body,K,LEntry,Exit,OldL,List,Vars_u,AbsInt,Id),
	each_exit_to_prime(Exit,AbsInt,Sg,Hv,Head,Sv,ExtraInfo,Prime1),
	widen_succ(AbsInt,TempPrime,Prime1,NewPrime),
	decide_flag(AbsInt,TempPrime,NewPrime,SgKey,Sg,Id,Proj,Flag).

do_r_cl(_,_,_,_,_,_,_,_,List,List,Prime,Prime,_).

widen_succ_off(AbsInt,Prime0,Prime1,LPrime):-
	current_pp_flag(multi_success,on), !,
	reduce_equivalent([Prime0,Prime1],AbsInt,LPrime).
widen_succ_off(AbsInt,Prime0,Prime1,Prime):-
	singleton(P0,Prime0),
	singleton(P1,Prime1),
	singleton(P,Prime), 
	compute_lub(AbsInt,[P0,P1],P).

reuse_entry(EntryInf,Vars_u,AbsInt,Sv,Sg,Hv,Head,K,Proj,Entry,ExtraInfo):-
	var(EntryInf), !,
	sort(Vars_u,Vars),
	ord_subtract(Vars,Hv,Fv),
	call_to_entry(AbsInt,Sv,Sg,Hv,Head,K,Fv,Proj,Entry,ExtraInfo),
	EntryInf = (Entry,ExtraInfo).
reuse_entry(EntryInf,_Vars_u,_AbsInt,_Sv,_Sg,_Hv,_Head,_K,_Proj,Entry,ExtraInfo):-
	EntryInf = (Entry,ExtraInfo).

decide_flag(AbsInt,TempPrime,NewPrime,_SgKey,_Sg,_Id,_Proj,_Flag):-
	abs_subset_(NewPrime,AbsInt,TempPrime), !.
decide_flag(_AbsInt,TempPrime,NewPrime,SgKey,Sg,Id,Proj,Flag):-
	Flag = notend,
	merge_(NewPrime,TempPrime,LPrime),
	current_fact(fixpoint(SgKey,Sg,_,_,Id,Fs),Ref),
	erase(Ref),
	asserta_fact(fixpoint(SgKey,Sg,Proj,LPrime,Id,Fs)),  
	current_fact(ch_id(Id,Num),Ref3),
	erase(Ref3),
	Num1 is Num+1,
	asserta_fact(ch_id(Id,Num1)).

merge_([NewPrime],_TempPrime,LPrime):- !, LPrime=[NewPrime].
merge_(NewPrime,TempPrime,LPrime):-
	merge(NewPrime,TempPrime,LPrime).

%------------------------------------------------------------------------%
% entry_to_exit(+,+,+,-,+,-,+,+,+)                                       %
%------------------------------------------------------------------------%

entry_to_exit((Sg,Rest),K,Call,Exit,OldList,NewList,Vars_u,AbsInt,NewN):- !,
	body_succ(Call,Sg,Succ,OldList,IntList,Vars_u,AbsInt,K,NewN,_),
	entry_to_exit(Rest,K,Succ,Exit,IntList,NewList,Vars_u,AbsInt,NewN).
entry_to_exit(true,_,Call,Call,List,List,_,_,_):- !.
entry_to_exit(Sg,Key,Call,Exit,OldList,NewList,Vars_u,AbsInt,NewN):- 
	body_succ(Call,Sg,Exit,OldList,NewList,Vars_u,AbsInt,Key,NewN,_),
	decide_memo(AbsInt,Key,NewN,no,Vars_u,Exit),!. 

body_succ(Call,Atom,Succ,List,List,HvFv_u,AbsInt,_ClId,ParentId,no):-
	bottom(Call), !,
	Succ = Call,
	Atom=g(Key,_Av,_I,_SgKey,_Sg),
	asserta_fact(memo_table(Key,AbsInt,ParentId,no,HvFv_u,Succ)).
body_succ(Call,Atom,Succ,List,NewList,HvFv_u,AbsInt,ClId,ParentId,Id):- 
	Atom=g(Key,Sv,Info,SgKey,Sg),
	body_succ_(Info,SgKey,Sg,Sv,HvFv_u,Call,Succ,List,NewList,AbsInt,
	           ClId,Key,ParentId,Id),
	decide_memo(AbsInt,Key,ParentId,Id,HvFv_u,Call).

body_succ_(Info,SgKey,Sg,Sv,HFv,Call,Succ,L,NewL,AbsInt,ClId,Key,PId,Id):-
	Info = [_|_], !,
	split_combined_domain(AbsInt,Call,Calls,Domains),
	map_body_succ(Info,SgKey,Sg,Sv,HFv,Calls,Succs,L,NewL,Domains,
	              ClId,Key,PId,Id),
	split_combined_domain(AbsInt,Succ,Succs,Domains).
body_succ_(Info,SgKey,Sg,Sv,HFv,Call,Succ,L,NewL,AbsInt,ClId,Key,PId,Id):-
	body_succ0(Info,SgKey,Sg,Sv,HFv,Call,Succ,L,NewL,AbsInt,
	           ClId,Key,PId,Id).

map_body_succ([],_SgKey,_Sg,_Sv,_HFv,[],[],L,L,[],_ClId,_Key,_PId,no).
map_body_succ([I|Info],SgKey,Sg,Sv,HFv,[Call|Calls],[Succ|Succs],L,NewL,
	      [AbsInt|Domains],ClId,Key,PId,Id):-
	body_succ0(I,SgKey,Sg,Sv,HFv,Call,Succ,L,_NewL,AbsInt,
	           ClId,Key,PId,_Id), !,
	map_body_succ(Info,SgKey,Sg,Sv,HFv,Calls,Succs,L,NewL,Domains,
	           ClId,Key,PId,Id).

body_succ0('$var',SgKey,Sg,_Sv_u,HvFv_u,Calls,Succs,List0,List,AbsInt,
	    ClId,F,_N,Id):-
        !,
	&( Calls=[Call],
	  concrete(AbsInt,Sg,Call,Concretes),
	  concretes_to_body(Concretes,SgKey,AbsInt,B)
	-> fixpoint_id(Id),
	   meta_call(B,HvFv_u,Calls,[],Succs,List0,List,AbsInt,ClId,Id,Ids),
	   assertz_fact(memo_call(F,Id,AbsInt,Concretes,Ids))
	 ; Id=no,
	   List=List0,
	   variable(F,ClId),
	   each_unknown_call(Calls,AbsInt,[Sg],Succs) % Sg is a variable
	&).
body_succ0('$meta'(T,B,_),SgKey,Sg,Sv_u,HvFv_u,Call,Succ,List0,List,AbsInt,
	    ClId,F,N,Id):-
        !,
	&( current_pp_flag(reuse_fixp_id,on) ->
	   &( Call=[C]
	     -> sort(Sv_u,Sv),
	        project(AbsInt,Sg,Sv,HvFv_u,C,Proj),
		fixpoint_id_reuse_prev(SgKey,AbsInt,Sg,Proj,Id)
              ; true
	   &)
	;
	    fixpoint_id(Id)
	&),
	meta_call(B,HvFv_u,Call,[],Exits,List0,List,AbsInt,ClId,Id,_Ids),
	&( body_succ_meta(T,AbsInt,Sv_u,HvFv_u,Call,Exits,Succ) ->
	   &( Call=[C] ->
	      sort(Sv_u,Sv),
	      project(AbsInt,Sg,Sv,HvFv_u,C,Proj),
	      each_project(Exits,AbsInt,Sg,Sv,HvFv_u,Prime),
	      asserta_fact(complete(SgKey,AbsInt,Sg,Proj,Prime,Id,[(F,N)]))
	    ; true
	   &)
	; % for the trusts, if any:
	  varset(Sg,Sv_r), % Sv_u contains extra vars (from meta-term)
                           % which will confuse apply_trusted
	  body_succ0(nr,SgKey,Sg,Sv_r,HvFv_u,Call,Succ,[],_List,AbsInt,
		        ClId,F,N,Id0), 
	  retract_fact(complete(SgKey,AbsInt,Sg,Proj,Prime,Id0,Ps)),
	  asserta_fact(complete(SgKey,AbsInt,Sg,Proj,Prime,Id,Ps))
	&).
body_succ0('$built'(T,Tg,Vs),SgKey,Sg,Sv_u,HvFv_u,Call,Succ,List0,List,AbsInt,
	    _ClId,F,N,Id):-
	!,
	Id=no,
	List=List0,
	sort(Sv_u,Sv),
	each_body_succ_builtin(Call,AbsInt,T,Tg,Vs,SgKey,Sg,Sv,HvFv_u,F,N,Succ).
body_succ0(RFlag,SgKey,Sg,Sv_u,HvFv_u,Call,Succ,List0,List,AbsInt,
	   ClId,F,N,Id):-
	sort(Sv_u,Sv),
	each_call_to_success(Call,RFlag,SgKey,Sg,Sv,HvFv_u,AbsInt,ClId,
                             Succ,List0,List,F,N,Id).

each_call_to_success([Call],RFlag,SgKey,Sg,Sv,HvFv_u,AbsInt,ClId,Succ,L0,L,
	             F,N,Id):-
	!,
	project(AbsInt,Sg,Sv,HvFv_u,Call,Proj),
	call_to_success(RFlag,SgKey,Call,Proj,Sg,Sv,AbsInt,ClId,Succ,L1,F,N,Id),

	merge(L1,L0,L).
each_call_to_success(LCall,RFlag,SgKey,Sg,Sv,HvFv_u,AbsInt,ClId,LSucc,L0,L,
                     F,N,Id):-
	each_call_to_success0(LCall,RFlag,SgKey,Sg,Sv,HvFv_u,AbsInt,ClId,
                              LSucc,L0,L,F,N,Id).

each_call_to_success0([],_Flag,_SgK,_Sg,_Sv,_HvFv,_AbsInt,_,[],L,L,_F,_N,_NN).
each_call_to_success0([Call|LCall],RFlag,SgKey,Sg,Sv,HvFv_u,AbsInt,ClId,
	              LSucc,L0,L,F,N,NewN):-
	project(AbsInt,Sg,Sv,HvFv_u,Call,Proj),
	call_to_success(RFlag,SgKey,Call,Proj,Sg,Sv,AbsInt,ClId,LSucc0,L1,F,N,_),
	merge(L0,L1,L2),
	append(LSucc0,LSucc1,LSucc),
	each_call_to_success0(LCall,RFlag,SgKey,Sg,Sv,HvFv_u,AbsInt,ClId,
	                      LSucc1,L2,L,F,N,NewN).

meta_call([],_HvFv_u,Call,[],Call,List,List,_AbsInt,_ClId,_Id,[]).
meta_call([Body|Bodies],HvFv_u,Call,Succ0,Succ,L0,List,AbsInt,ClId,Id,Ids):-
	meta_call_([Body|Bodies],HvFv_u,Call,Succ0,Succ,L0,List,AbsInt,ClId,Id,Ids).

meta_call_([Body|Bodies],HvFv_u,Call,Succ0,Succ,L0,List,AbsInt,ClId,Id,Ids):-
	meta_call_body(Body,ClId,Call,Succ1,L0,L1,HvFv_u,AbsInt,Id,Ids0),
	widen_succ(AbsInt,Succ0,Succ1,Succ2),
	append(Succ0,Succ1,Succ2),
	append(Ids0,Ids1,Ids),
	meta_call_(Bodies,HvFv_u,Call,Succ2,Succ,L1,List,AbsInt,ClId,Id,Ids1).
meta_call_([],_HvFv_u,_Call,Succ,Succ,List,List,_AbsInt,_ClId,_Id,[]).

meta_call_body((Sg,Rest),K,Call,Exit,OldList,NewList,Vars_u,AbsInt,PId,CIds):-
	!,
	CIds=[Id|Ids],
	body_succ(Call,Sg,Succ,OldList,IntList,Vars_u,AbsInt,K,PId,Id),
	meta_call_body(Rest,K,Succ,Exit,IntList,NewList,Vars_u,AbsInt,PId,Ids).
meta_call_body(true,_,Call,Call,List,List,_,_,_,[no]):- !.
meta_call_body(Sg,Key,Call,Exit,OldList,NewList,Vars_u,AbsInt,PId,[Id]):- 
	body_succ(Call,Sg,Exit,OldList,NewList,Vars_u,AbsInt,Key,PId,Id).

concretes_to_body([],_SgKey,_AbsInt,[]).
concretes_to_body([Sg|Sgs],SgKey,AbsInt,[B|Bs]):-
	body_info0(Sg:SgKey,[],AbsInt,B),
	concretes_to_body(Sgs,SgKey,AbsInt,Bs).

%------------------------------------------------------------------------%
% query(+,+,+,+,+,+,+,-)                                                 %
%------------------------------------------------------------------------%

:- doc(query(AbsInt,QKey,Query,Qv,RFlag,N,Call,Succ),
	"The success pattern of @var{Query} with @var{Call} is
         @var{Succ} in the analysis domain @var{AbsInt}. The predicate
         called is identified by @var{QKey}, and @var{RFlag} says if it
         is recursive or not. The goal @var{Query} has variables @var{Qv},
         and the call pattern is uniquely identified by @var{N}.").

query(AbsInt,QKey,Query,Qv,RFlag,N,Call,Succ) :-
	project(AbsInt,Query,Qv,Qv,Call,Proj),
	call_to_success(RFlag,QKey,Call,Proj,Query,Qv,AbsInt,0,Succ,_,N,0,Id), !,
	approx_to_completes(AbsInt).

query(_AbsInt,_QKey,_Query,_Qv,_RFlag,_N,_Call,_Succ):-
	% should never happen, but...
	error_message("SOMETHING HAS FAILED!").
\end{lstlisting}

\end{document}